\documentclass[12pt]{article}
\usepackage[dvips]{graphicx}
\usepackage{epsfig} 
\psfigdriver{dvips}
\begin{document}
\setcounter{page}{1}
\renewcommand{\thefootnote}{\fnsymbol{footnote}}
\renewcommand{\theenumi}{(\arabic{enumi})}
\renewcommand{\thefigure}{\arabic{figure}}
\renewcommand{\theequation}{\arabic{equation}}
\renewcommand{\arraystretch}{1.3}
\def\bea{\begin{eqnarray}}
\def\eea{\end{eqnarray}}
\def\bseq{\begin{subequations}}
\def\eseq{\end{subequations}}
\def\nn{\nonumber}
\def\dfrac{\displaystyle\frac}
\def\numt#1#2{#1 \times 10^{#2}}
\def\etal{{\it et al.}}
\def\eg{{\it e.g.~}}
\def\bs{\bigskip}

\def\PR#1#2#3{Phys. Rev. {\bf #1}, #2 (#3)}
\def\PRL#1#2#3{Phys. Rev. Lett. {\bf #1}, #2 (#3)}
\def\PL#1#2#3{Phys. Lett. {\bf #1}, #2 (#3)}
\def\NL#1#2#3{Nucl. Phys. {\bf #1}, #2 (#3)}
\def\NP#1#2#3{Nucl. Phys. {\bf #1}, #2 (#3)}
\def\PREP#1#2#3{Phys. Report {\bf #1}, #2 (#3)}
\def\Mod#1#2#3{Mod. Phys. Lett. {\bf #1}, #2 (#3)}
\def\PTP#1#2#3{Prog. Theor. Phys. {\bf #1}, #2 (#3)}
\def\EPJ#1#2#3{Eur. Phys. J. {\bf #1}, #2 (#3)}
\def\MPLA#1#2#3{Mod. Phys. Lett. {\bf A#1} (19#2) #3}
\def\PRD#1#2#3{Phys. Rev. {\bf D#1} (#2) #3}
\def\NPB#1#2#3{Nucl. Phys. {\bf B#1} (#2) #3}
\def\ZPC#1#2#3{Z. Phys. {\bf C#1} (#2) #3}
\def\EPJC#1#2#3{Eur. Phys. J. {\bf C#1} (#2) #3}
\def\PLB#1#2#3{Phys. Lett. {\bf B#1} (#2) #3}
\def\PRep#1#2#3{Phys. Rep. {\bf #1} (#2) #3}

\def\eqref#1{eq.(\ref{eqn:#1})}
\def\eqlab#1{\label{eqn:#1}}
\def\Fgref#1{Fig.\ref{fig:#1}}
\def\Fglab#1{\label{fig:#1}}

\def\ov{\overline}
\def\l{\left}
\def\r{\right}
\def\gsim{~{\rlap{\lower 3.5pt\hbox{$\mathchar\sim$}}\raise 1pt\hbox{$>$}}\,}
\def\lsim{~{\rlap{\lower 3.5pt\hbox{$\mathchar\sim$}}\raise 1pt\hbox{$<$}}\,}
\def\bpm{~{\rlap{\lower 5pt\hbox{$~-$}}\raise 3pt\hbox{\small$(+)$}}\,}
\def\bbpm{{\lower 5pt\hbox{$~-$}}\raise 3pt\hbox{\small$\hspace{-13pt}(+)$}}
\makeatletter
\newtoks\@stequation
\def\subequations{\refstepcounter{equation}%
  \edef\@savedequation{\the\c@equation}%
  \@stequation=\expandafter{\theequation}
  \edef\@savedtheequation{\the\@stequation}
  \edef\oldtheequation{\theequation}%
  \setcounter{equation}{0}%
  \def\theequation{\oldtheequation\alph{equation}}}
\def\endsubequations{%
  \ifnum\c@equation < 2 \@warning{Only \the\c@equation\space subequation
    used in equation \@savedequation}\fi
  \setcounter{equation}{\@savedequation}%
  \@stequation=\expandafter{\@savedtheequation}%
  \edef\theequation{\the\@stequation}%
  \global\@ignoretrue}
\def\eqnarray{\stepcounter{equation}\let\@currentlabel\theequation
\global\@eqnswtrue\m@th
\global\@eqcnt\z@\tabskip\@centering\let\\\@eqncr
$$\halign to\displaywidth\bgroup\@eqnsel\hskip\@centering
     $\displaystyle\tabskip\z@{##}$&\global\@eqcnt\@ne
      \hfil$\;{##}\;$\hfil
     &\global\@eqcnt\tw@ $\displaystyle\tabskip\z@{##}$\hfil
   \tabskip\@centering&\llap{##}\tabskip\z@\cr}
\makeatother
\def\cU#1#2{U_{#1}^{#2}}
\def\satms#1{\sin^2\theta_{_{\rm ATM}}^{#1}}
\def\satmw#1{\sin^22\theta_{_{\rm ATM}}^{#1}}
\def\ssun#1{\sin^22\theta_{_{\rm SOL}}^{#1}}
\def\srct#1{\sin^22\theta_{_{\rm RCT}}^{#1}}
\def\matm#1{\delta m^{2~#1}_{_{\rm ATM}}}
\def\msun#1{\delta m^{2~#1}_{_{\rm SOL}}}
\def\dmns#1{\delta_{_{\rm MNS}}^{#1}}
\begin{titlepage}
\thispagestyle{empty}
\begin{flushright}
\begin{tabular}{l}
{HIP-2003-57/TH} \\
{KEK-TH-926}\\
{KIAS-P03081}\\
{VPI-IPPAP-03-18}\\
\\
\end{tabular}
\end{flushright}
\baselineskip 24pt 
\begin{center}
{\Large\bf
Lifting degeneracies in the oscillation parameters 
by a neutrino factory
}
\vspace{5mm}

\baselineskip 18pt 
\renewcommand{\thefootnote}{\fnsymbol{footnote}}
\setcounter{footnote}{0}

{Mayumi Aoki$^{1,2}$\footnote{{mayumi.aoki@kek.jp}},
 Kaoru Hagiwara$^{2}$,
 and
 Naotoshi Okamura$^{3,4}$\footnote{{nokamura@kias.re.kr}}
}\\
\bs
\bs
{\small \it $^{1}$Helsinki Institute of Physics, P.O.Box 64
FIN-00014, University of Helsinki, Finland}\\
{\small \it $^{2}$Theory Group, KEK, Tsukuba, Ibaraki 305-0801, Japan}\\
{\small \it $^{3}$IPPAP, Physics Department, Virginia Tech. Blacksburg, VA
24061, USA}\\
{\small \it $^{4}$Korea Institute for Advanced Study, 207-43 Cheongnyangni 2dong}\\
{\small \it Dongdaemun-gu, Seoul 130-722, Republic of Korea}
\\
\end{center}
\begin{abstract}

We study the potential of a very long baseline neutrino oscillation 
experiment with a neutrino factory and a large segmented
water-$\check{\rm C}$erenkov calorimeter
detector in resolving the degeneracies 
in the neutrino oscillation parameters; 
the sign of the larger mass-squared difference $\delta m^2_{13}$, 
the sign of $|U_{_{\mu 3}}|^2 (\equiv \sin^2\theta_{_{\rm ATM}})-1/2$, 
and a possible two-fold ambiguity in the determination of the 
CP phase $\delta_{_{\rm MNS}}$.
We find that the above problems can be resolved even if the particle
charges are not measured.
The following results are obtained in our exploratory study
for a neutrino factory which 
delivers $10^{21}$ decaying $\mu^+$ and $\mu^-$ at 10 GeV and 
a 100 kton detector which is placed 2,100 km away and is capable of
measuring the event energy and distinguishing $e^\pm$ from $\mu^\pm$,
but not their charges.  
The sign of $\delta m^2_{13}$ can be determined for 
$4|U_{e3}|^2(1-|U_{e3}|^2)\equiv \sin^22\theta_{_{\rm RCT}}\gsim 0.008$.
That of $\sin^2\theta_{_{\rm ATM}}-1/2$ can be resolved 
for $\sin^22\theta_{_{\rm ATM}}=0.96$ 
when $\sin^22\theta_{_{\rm RCT}}\gsim 0.06$.
The CP-violating phase $\delta_{_{\rm MNS}}$ can be uniquely constrained for 
$\sin^22\theta_{_{\rm RCT}}\gsim 0.02$ if its true value is around 
$90^\circ$ or $270^\circ$, while 
it can be constrained for $\sin^22\theta_{_{\rm RCT}}\gsim 0.03$
if its true value is around $0^\circ$ or $180^\circ$. 

\end{abstract}

\bs
\bs
{
 \small 
 \begin{flushleft}
 \end{flushleft}
}
\begin{flushleft}
\end{flushleft}
\end{titlepage}
\newpage
\setcounter{page}{1}
\renewcommand{\thepage}{\arabic{page}}
\baselineskip 18pt 
\renewcommand{\thefootnote}{\fnsymbol{footnote}}
\setcounter{footnote}{0}

Atmospheric neutrino observation at Super-Kamiokande (SK) 
\cite{atm_tau} and the K2K experiment \cite{K2K} 
established $\nu_\mu^{}$ oscillation 
with a close-to-maximal mixing $\satmw{} > 0.9$ 
and $\matm{}\sim (1.3 - 3.0)\times 10^{-3}$ eV$^2$. 
The MSW large-mixing-angle (LMA) solution \cite{wolf,MSW} 
of the solar-neutrino deficit problem \cite{solar,SK_sun,SNO} has
been established by KamLAND \cite{KamLAND} and
by the improved measurement by SNO \cite{SNO}.
The best fit values are
$\ssun{}= 0.82$ and $\msun{}= 7.1\times 10^{-5}$ eV$^2$.
The CHOOZ \cite{CHOOZ} and Palo Verde \cite{PaloVarde} reactor experiments 
give upper bounds on the third mixing angle of the three-neutrino model; 
$\srct{}\lsim 0.1~(0.2)$ for
$\matm{} \sim 3.0~(2.0)\times 10^{-3}$ eV$^2$.

In the three-neutrino model, the present neutrino oscillation
experiments constrain
$U_{e2}$, $U_{e3}$, and $U_{\mu 3}$ matrix elements of 
the lepton-flavor mixing matrix
$U_{_{\rm MNS}}$  (Maki-Nakagawa-Sakata (MNS) \cite{MNS}) :
\bseq
\begin{eqnarray}
|U_{\mu 3}|^2&\equiv&
\satms{}
=\left(1 \pm \sqrt{ 1-\sin^2 2\theta_{_{\rm ATM}}}\right)
\scalebox{1.3}{$/^{^{}}$}2\,,
\eqlab{def_Um3}\\
|U_{e 2}|^2 &=&
\left(1 - { \l|U_{e3}\r|^2} -
\sqrt{\l(1 - { \l|U_{e3}\r|^2}\r)^2
-\sin^2 2\theta_{_{\rm SOL}}}\right)
\scalebox{1.7}{$/^{^{}}$}2\,,
\eqlab{def_Ue2} \\
{{\l|U_{e 3}\r|^2}} &=&
\left(1 - \sqrt{ 1-\sin^2 2\theta_{_{\rm RCT}}}\right)
\scalebox{1.3}{$/^{^{}}$}2\,, 
\eqlab{def_Ue3}
\end{eqnarray}
\eqlab{def_U}
\eseq
\hspace*{-1.0em}
and
the mass-squared differences,
$
 \delta m^2_{_{\rm SOL}} = \l|\delta {m}^2_{12}\r| \ll
\l|\delta {m}^2_{13}\r|  = \delta m^2_{_{\rm ATM}}
\eqlab{relation_MSD}\,,
$
where $\delta {m}^2_{ij}\equiv m^2_j-m^2_i$.
The matrix elements
$U_{e2}$ and $U_{\mu 3}$ 
are taken to be real and non-negative
while $U_{e3}$ is a complex number in our convention \cite{HO,H2B}.
The CP-violating phase $\dmns{}=-arg(U_{e3})$, is unconstrained.
Note that
the solution for $U_{e2}$ follows from our convention
$U_{e1}>U_{e2}$ \cite{H2B}, which defines the
mass-eigenstate $\nu_1^{}$.

The present experiments allow
several degeneracies 
in the neutrino oscillation parameters.
First is the sign of the larger mass-squared difference,
$\delta m^2_{13} \equiv m_3^2-m^2_1=\pm \matm{}$.
The sign of the smaller mass-squared difference is determined as
$\delta m^2_{12} \equiv m^2_2 - m^2_1 =\msun{}$ by the MSW matter
effects in the sun.
The mass hierarchy $\delta m^2_{13}=\matm{}$ is called `normal' and
$\delta m^2_{13}=-\matm{}$ is called `inverted'. 
The second degeneracy is in the mixing angle $\theta_{_{\rm ATM}}$.
Although the present experiments constrain $\satmw{}\simeq 1$,
$\satms{}$ has two-fold ambiguity, $\satms{}-1/2 = \pm\sqrt{1-\satmw{}}/2$.
Furthermore, there can appear two-fold ambiguity in the future
determination of the $\dmns{}$ phase, between $\dmns{}$ and 
$180^\circ -\dmns{}$.
Implications of the above degeneracy problems and their partial
resolutions have been proposed in \cite{NeutrinoFactory}.

The proposed long-baseline (LBL) neutrino-oscillation 
experiments \cite{MINOS,JHF2SK}
will achieve
the precision measurements of $\delta m^2_{_{\rm ATM}}$ and 
$\sin^22\theta_{_{\rm ATM}}$ 
by using conventional neutrino beams,
which are made from decays of $\pi$ and $K$ that are produced 
by high-energy proton beams.
However, lifting the above eight-fold degeneracy in the neutrino
oscillation parameters could remain as the target for the next
generation neutrino experiments.

In Ref.\cite{H2H}, we studied the possibility of measuring $\dmns{}$
in the LBL experiment with J-PARC
(Japan Proton Accelerator Complex) \cite{JPARC} at Tokai Village
and a Megaton-level
water-$\check {\rm C}$erenkov detector, 
Hyper-Kamiokande (HK) \cite{Hyper-K}.
The distance between Tokai Village and HK is about 295 km.
There we found that it is relatively easy to distinguish between
$\dmns{}=90^\circ$ and $270^\circ$ cases, but that it is difficult to
distinguish between $\dmns{}=0^\circ$ and $180^\circ$ cases.
Those studies, however, assumed that the neutrino mass hierarchy is
known to be normal $(\delta m^2_{13}=\matm{})$ and also we assumed
$\satms{}=0.5$ as an input.

In \Fgref{H2H}(a),
we show the allowed regions of Tokai-to-HK experiment 
when $\dmns{}=90^\circ$ and the normal hierarchy is assumed.
CP-violation can be established at better than $3\sigma$ level
if the neutrino mass hierarchy is known.
However, the region encircled by thin lines are allowed if the
inverted hierarchy $(\delta m^2_{13}=-\matm{})$ is assumed in the
analysis of the same data.
CP conservation is preferred in the latter case.
In \Fgref{H2H}(b),
we show the allowed region of the same experiment when
$\satms{\rm true} = 0.45,~0.35$.
Mirror solutions at around $\satms{\rm mirror} = 0.55,~0.65$, respectively,
are clearly seen\footnote{{%
The results in \Fgref{H2H} are obtained by using the off-axis beams
\cite{JHF2SK}.
Details of the J-PARC-to-HK
analyses with the off-axis beams
will be reported elsewhere \cite{NBBvsOAB}.
}}.

\begin{figure}[ht]
\epsfig{file=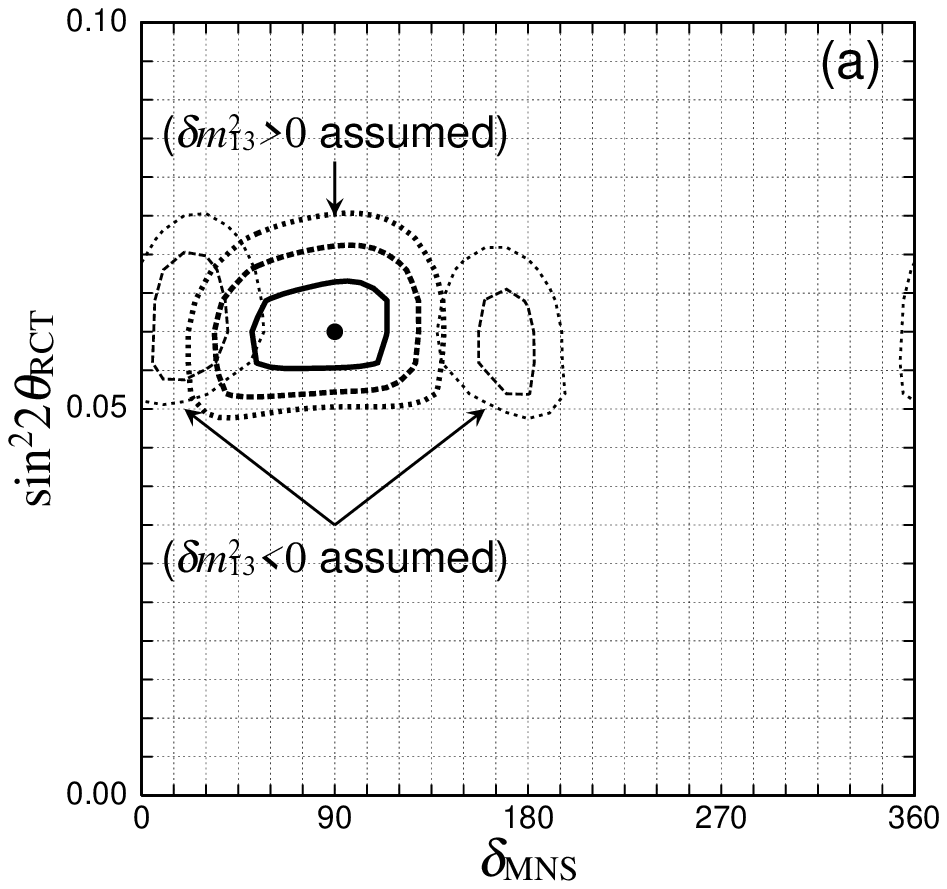,width=8cm,angle=0}
\epsfig{file=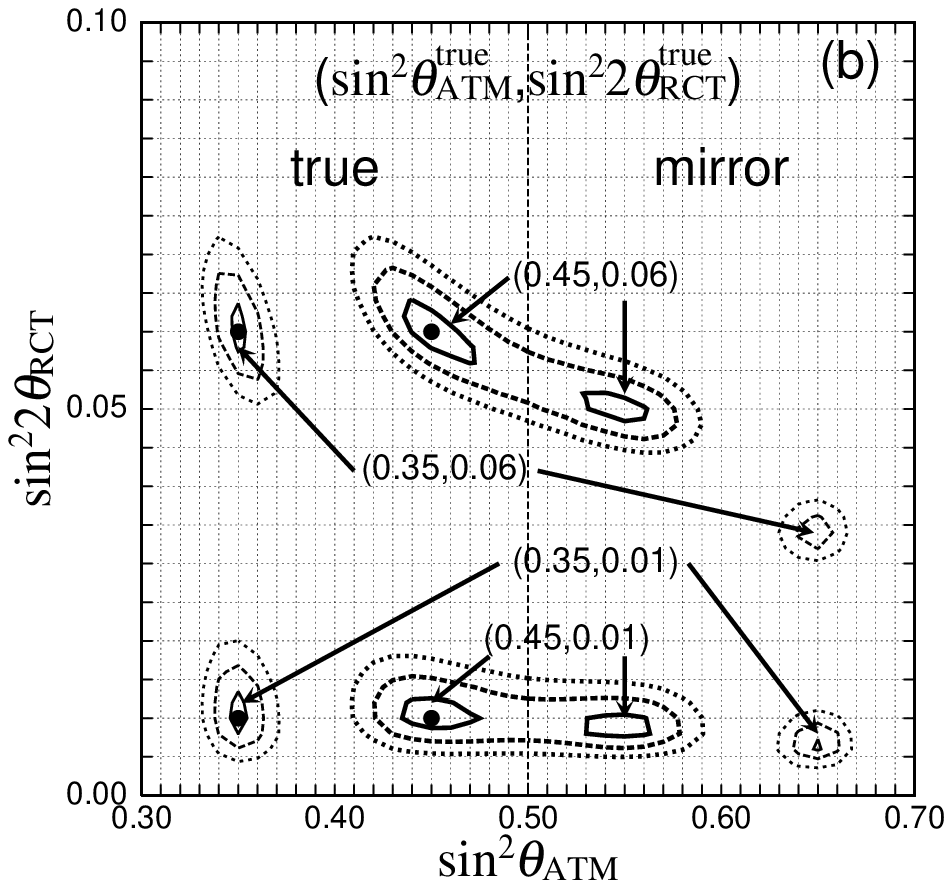,width=8cm,angle=0}
\caption{%
The regions allowed by future LBL experiment between 
J-PARC at Tokai and HK
in the plane of 
$\delta_{_{\rm MNS}}$ and $\sin^22\theta_{_{\rm RCT}}$ (a),
and $\sin^2\theta_{_{\rm ATM}}$ and $\sin^22\theta_{_{\rm RCT}}$ (b).
The assumed experimental conditions are  
0.8 Mton$\cdot$year for $\nu_\mu$ OAB($2^\circ$),  
2 Mton$\cdot$year for $\nu_\mu$ OAB($3^\circ$), and
3.2 Mton$\cdot$year for $\ov \nu_\mu$ OAB($2^\circ$).
$\chi^2$ fittings are performed by assuming the normal and the 
inverted hierarchies in (a) and the normal hierarchy only in (b). 
}
\Fglab{H2H}
\end{figure}

In this paper, we examine the possibility
of a very long-baseline (VLBL)
neutrino oscillation experiment with a neutrino factory at
J-PARC in resolving
the neutrino mass hierarchy, 
the sign of $\sin^2\theta_{_{\rm ATM}}-1/2$,
and the degeneracy in the CP phase $\delta_{_{\rm MNS}}$.  
The possibility of a neutrino factory at J-PARC in Tokai
is studied in Ref.\cite{NF_Japan}.
For the detector, we assume a 100 kton-level segmented
water-$\check {\rm C}$erenkov calorimeter detector at 
$L=2,100$ km away from Tokai.
The distance is approximately that between Tokai and Beijing,
where strong interests in constructing a large 
water-$\check {\rm C}$erenkov detector
BAND
(Beijing Astrophysics and Neutrino Detector) have been expressed \cite{BAND}. 
The BAND detector has a good capability for detecting $\nu_e$ 
charged current (CC) and
$\nu_\mu$ CC events with calorimetric energy measurement
\cite{BAND,BAND2}.
In this analysis, we assume that the detector is capable of measuring
the event energy and distinguishing $e^\pm$ from $\mu^\pm$, but 
we do not require its charge identification capability.

The physics prospects of VLBL oscillation 
experiments with a neutrino factory 
has been studied in the past by assuming that the
detector can identify charges \cite{NeutrinoFactory}, 
and hence the possibility 
of charge identification at a large
water-$\check {\rm C}$erenkov detector
has been investigated \cite{BAND2}.
We would like to show in this paper that even if a detector is 
charge blind it 
could achieve all the goals.

In the neutrino factory, neutrinos are produced 
from the decay of high energy muons, 
$\mu^+ \to \ov\nu_\mu \nu_e e^+$ or $\mu^- \to \nu_\mu \ov\nu_e e^- $.
The same amount of
$\ov\nu_\mu$ and $\nu_e$ ($\nu_\mu$ and $\ov\nu_e$ ) are 
contained in the stored $\mu^+$ ($\mu^-$) beam.
One of the most significant qualities of the neutrino factory is the
well knowledge of neutrino fluxes.
Assuming very relativistic muons,
the $\ov \nu_\mu$ and $\nu_e$ 
($\nu_\mu$ and $\ov\nu_e$ ) fluxes 
from $\mu^+ (\mu^-)$ beam
are expressed as
\bseq
\bea
\Phi_{\ov \nu_\mu (\nu_\mu)} &=&
\gamma^2\frac{n_\mu}{\pi L^2}2y^2
\left\{(3-2y)\mp 
P_\mu(1-2y)\right\} 
\,,
\eqlab{flux_mu} \\
\Phi_{\nu_e (\ov \nu_e)}
&=&\gamma^2\frac{n_\mu}{\pi L^2}
12y^2
\left\{(1-y) \mp P_\mu(1-y)\right\} 
\,,
\eqlab{flux_e} 
\eea
\eseq
where $\gamma=E_\mu/m_\mu$,
$y=E_\nu/E_\mu$ with the energy of the decaying muon $E_\mu$,
$P_\mu$ is the average muon polarization
($P_\mu=1$ for right-handed and $P_\mu=-1$ for left-handed $\mu^\pm$), 
and
$n_\mu$ is the number of the decaying muons.
The upper (lower) sign should be taken for $\mu^+ (\mu^-)$ beam.

The $e$-like signal from $\mu^+$ beam, $N_e(\mu^+)$,
is given by the sum of
$e^+$ from the $\ov\nu_\mu \to \ov\nu_e$ appearance mode and 
$e^-$ from the $\nu_e \to \nu_e $ survival mode,
whereas
the $\mu$-like signal, $N_{\mu}(\mu^+)$,  
is the sum of
$\mu^+$ from the $\ov\nu_\mu \to \ov\nu_\mu$ survival mode and 
$\mu^-$ from the $\nu_e \to \nu_\mu $ appearance mode.
The signals from the $\mu^-$ beam,
$N_e(\mu^-)$ and  $N_{\mu}(\mu^-)$,
are obtained in the same way;
\bseq
\bea
N_e(\mu^+)~~ &:& ~~\nu_e \to\nu_e ~~+~~\ov\nu_\mu \to \ov\nu_e\,,
\\
N_\mu(\mu^+)~~ &:& ~~\ov\nu_\mu \to \ov\nu_\mu~~ +~~\nu_e \to\nu_\mu
\,,\\
N_e(\mu^-)~~ &:& ~~\ov\nu_e \to\ov\nu_e ~~+~~\nu_\mu \to \nu_e
\,,\\
N_\mu(\mu^-)~~ &:& ~~\nu_\mu \to \nu_\mu ~~+ ~~\ov\nu_e \to\ov\nu_\mu
\,.
\eea
\eseq
The signals
in the {\it i}-th energy bin,
$N^i_l(\mu^+)$ and  $N^i_l(\mu^-)$ ($l=e$ or $\mu$),
are then calculated as 
\bseq
\eqlab{N_pm}
\begin{eqnarray}
 N^i_l(\mu^+)&=&M~N_A^{} 
{\displaystyle \int}_{E_i}^{E_i+\delta E}
d E_\nu 
\left\{ \Phi_{\ov \nu_\mu} \cdot
  P_{\ov \nu_\mu^{} \to \ov \nu_{l}^{}} \cdot
\sigma^{\rm CC}_{\ov \nu_l}
+ \Phi_{\nu_e}
\cdot  P_{\nu_e^{} \to \nu_{l}^{}}
\cdot \sigma^{\rm CC}_{\nu_l}
\right\}\,,
 \eqlab{N_mu+}  \\
 N^i_l(\mu^-)&=&M~N_A^{}
{\displaystyle \int}_{E_i}^{E_i+\delta E}
d E_\nu
\left\{ \Phi_{\nu_\mu}
\cdot  P_{\nu_\mu^{} \to \nu_{l}^{}}
\cdot \sigma^{\rm CC}_{\nu_l}
+ \Phi_{\ov \nu_e}
\cdot  P_{\ov \nu_e^{} \to \ov \nu_{l}^{}}
\cdot \sigma^{\rm CC}_{\ov \nu_l}
\right\}\,,
\eqlab{N_mu-}
\end{eqnarray}
\eseq
\noindent
respectively,
where 
$P_{\nu_\alpha \to \nu_\beta}$ and $P_{\ov \nu_\alpha \to \ov \nu_\beta}$
are the neutrino and anti-neutrino
oscillation probabilities inside the earth matter,
$\delta E$ is the width of the energy bin, 
$M$ is the mass of the detector,
$N_A=\numt{6.017}{23}$ is the Avogadro number,
and
$\sigma_{\nu_l}^{\rm CC}$ and 
$\sigma_{\ov \nu_l}^{\rm CC}$ are the $\nu_l$ and $\ov \nu_l$
CC cross sections of water target \cite{cross}.

We show 
the dependences of
the signals, $N_l^i(\mu^+)$ and $N_l^i(\mu^-)$,
on the mass hierarchy and $\sin^2\theta_{_{\rm ATM}}$
in \Fgref{signal}
with $10^{21}$ of unpolarized $\mu^+$ and $\mu^-$ decays,
respectively, at 10 GeV.
The left figures are the signals from the $\mu^+$ beam
and the right ones are from the $\mu^-$ beam.
The width of the bin is taken as $\delta E=$1 GeV for $E_\nu > 2$ GeV. 
In \Fgref{signal}, 
the signals for the normal and the inverted mass hierarchies are 
shown by thick and thin lines, respectively.
We choose three values of $\sin^2\theta_{_{\rm ATM}}=$ 
0.5 (solid lines),
0.35 (dashed lines), and 0.65 (dot-dashed lines),
where the last two values 
give $\sin^22\theta_{_{\rm ATM}}=0.91$. 
The other parameters with
the constant matter density $\rho$  
are  
\bea
\eqlab{signal_bin}
\delta m^{2}_{_{\rm ATM}}&=&3 \times10^{-3} ~{\rm eV}^2
\,, ~~~~~~~~~~~~
\delta m^{2}_{_{\rm SOL}}=7\times10^{-5} ~{\rm eV}^2
\,,
\nn \\
\sin^22\theta_{_{\rm SOL}}&=&0.85
\,,~~~~~~~~~~~~~~~~~~~~~~~~~
\delta_{_{\rm MNS}}=0^\circ
\,, 
\\
\sin^22\theta_{_{\rm RCT}}&=&0.06
\,,~~~~~~~~~~~~~~~~~~~~~~~~~
\rho=3 ~{\rm g/cm}^3\,.  
\nn
\eea
\Fgref{signal_rct} shows the dependences of the expected number of
events,
$N_l^i(\mu^+)$ and $N_l^i(\mu^-)$,
on $\sin^22\theta_{_{\rm RCT}}$
assuming the normal hierarchy for the same experimental setup as 
in \Fgref{signal}.
We take three values of 
$\sin^22\theta_{_{\rm RCT}}=0.1$ (dashed lines), 
0.06 (solid lines), and 0 (dotted lines).
The other parameters are 
taken as \eqref{signal_bin} 
and $\sin^2\theta_{_{\rm ATM}}=0.5$.
In \Fgref{signal_rct}, 
we also show the number of background events by
shaded bars, where
In this study, we take account of the events coming from   
$\tau$ pure-leptonic-decays $N_l(\mu^\pm,\tau\to l)$, 
neutral-current (NC) events, $N_l(\mu^\pm,{\rm NC})$,
and
$\tau$ hadronic-decays, $N_l(\mu^\pm,\tau \to {\rm had})$.
The last two processes 
can contribute to $e$-like events,  
where produced $\pi^0$'s mimic the electron
shower in the detector. 

\begin{figure}[th]
\begin{center}
\epsfig{file=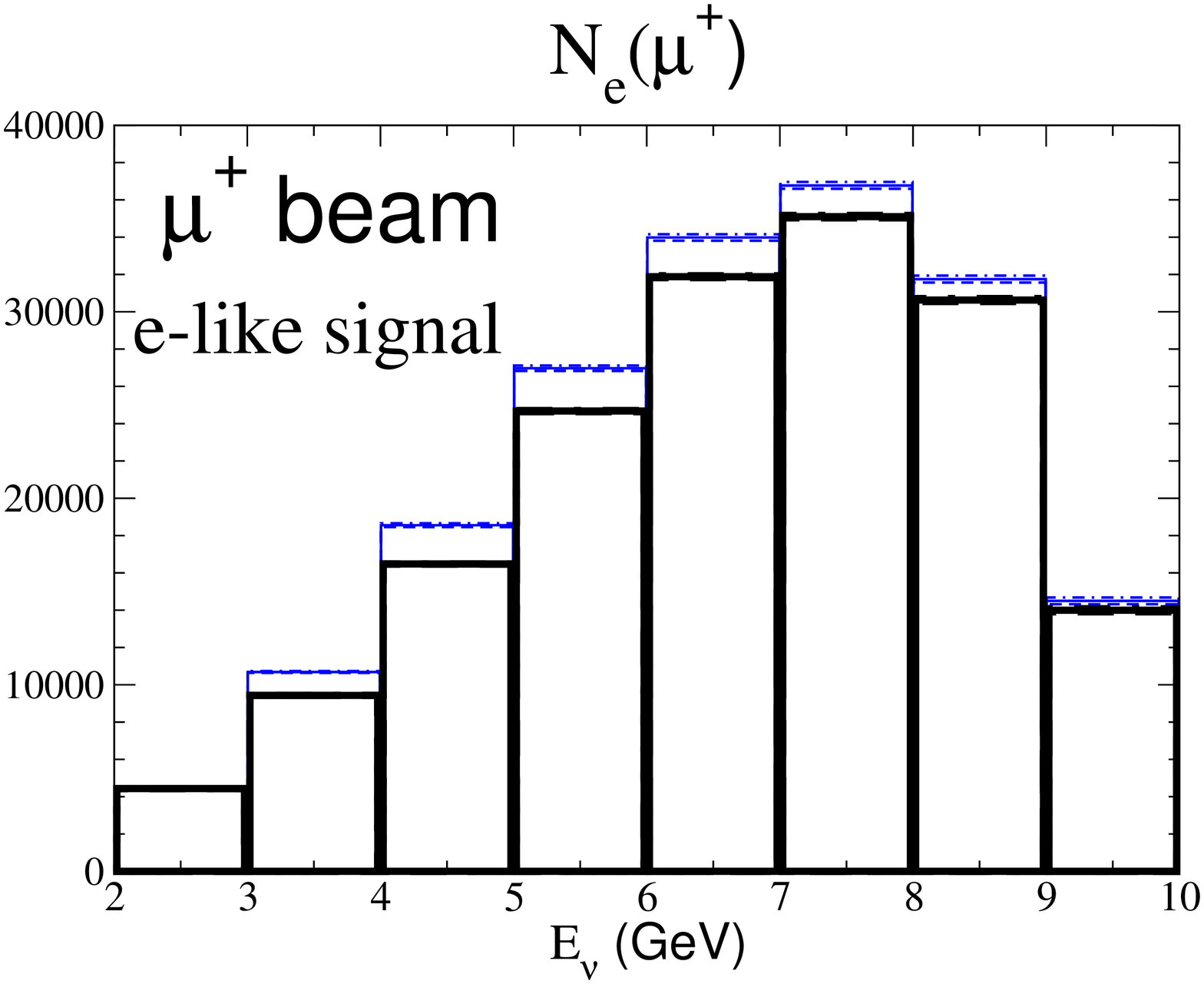,width=6cm,angle=-90}
\epsfig{file=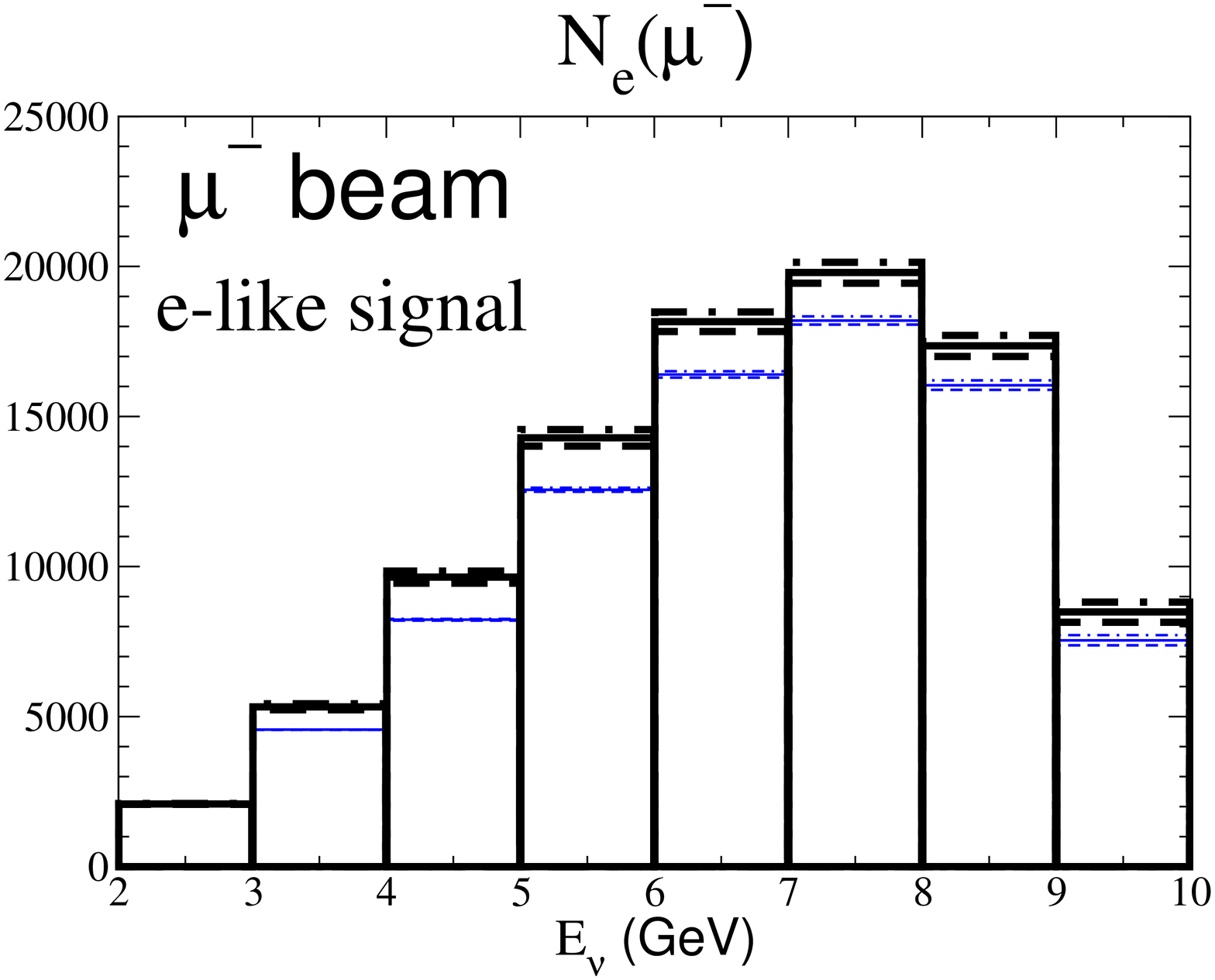,width=6cm,angle=-90}
\epsfig{file=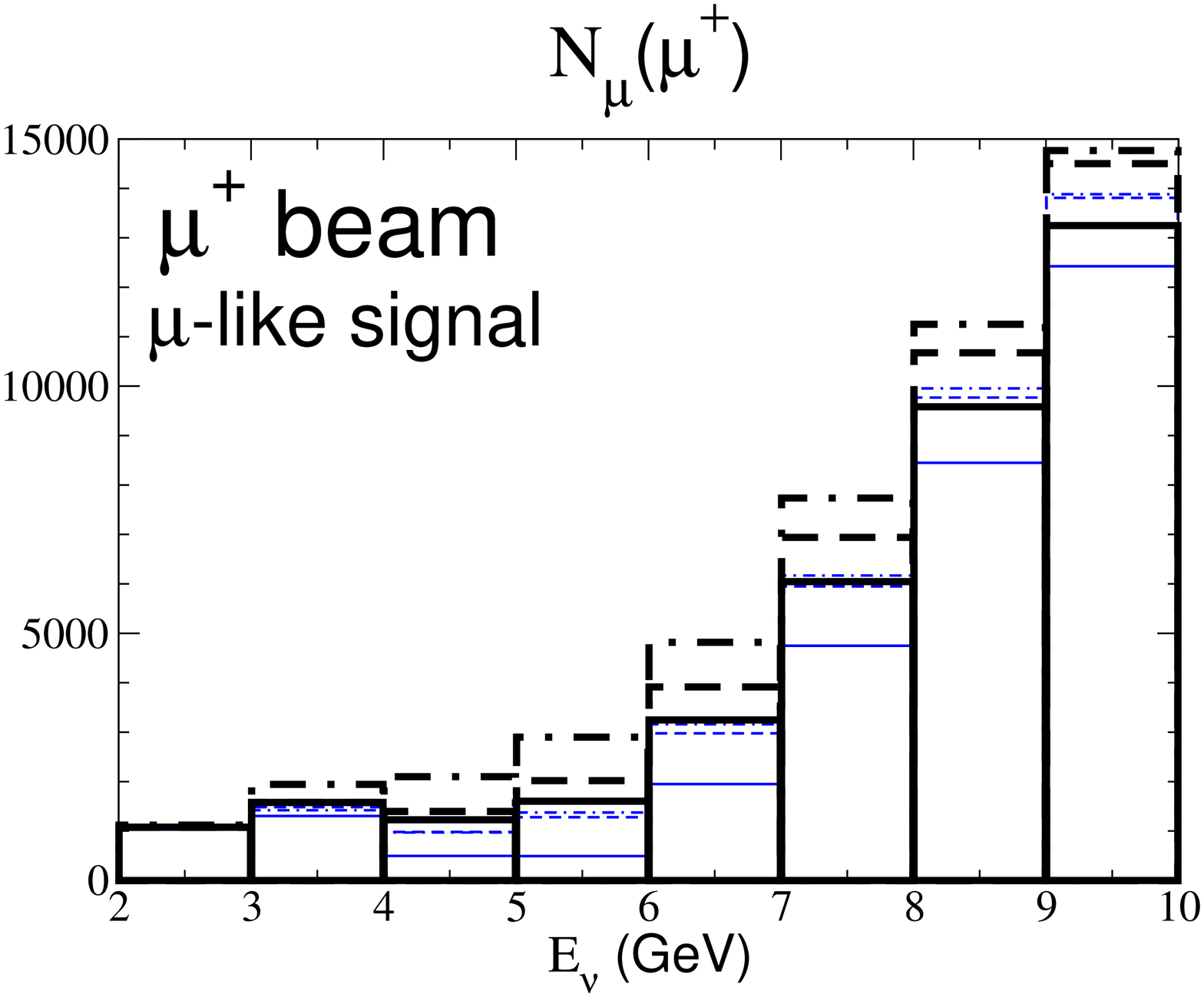,width=6cm,angle=-90}
\epsfig{file=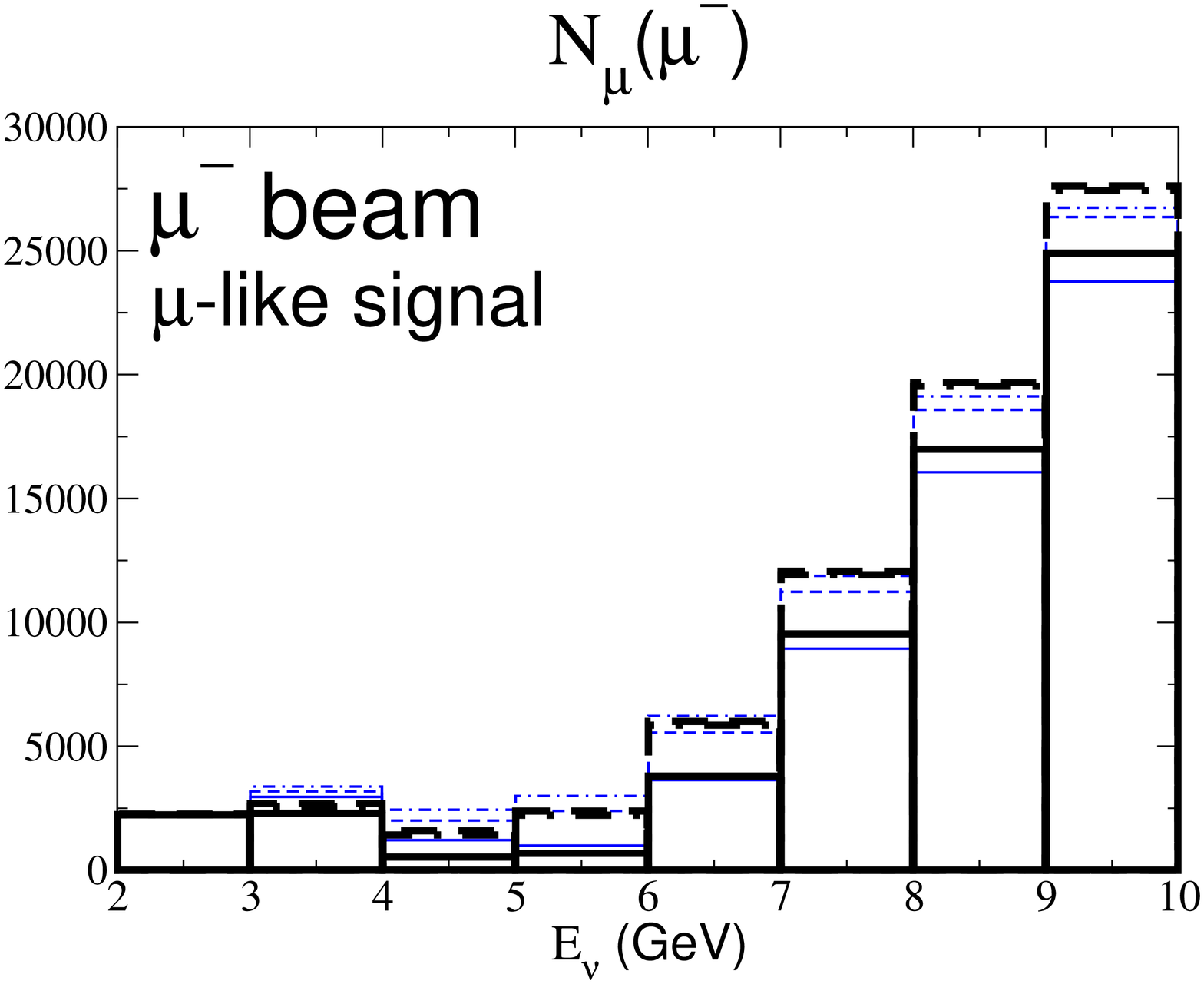,width=6cm,angle=-90}
\caption{Number of $e$- and $\mu$-like signals 
for $10^{21}$ decaying $\mu^+$ (left) and $\mu^-$ (right) each
at 10 GeV
for the normal-hierarchy (thick lines) and the inverted hierarchy (thin lines) 
with three values of $\sin^2\theta_{_{\rm ATM}}=$0.5 (solid lines),
0.35 (dashed lines), and 0.65 (dot-dashed lines).
The other parameters are taken as in \eqref{signal_bin}.
}
\Fglab{signal}
\end{center}
\end{figure}

\begin{figure}[ht]
\begin{center}
\epsfig{file=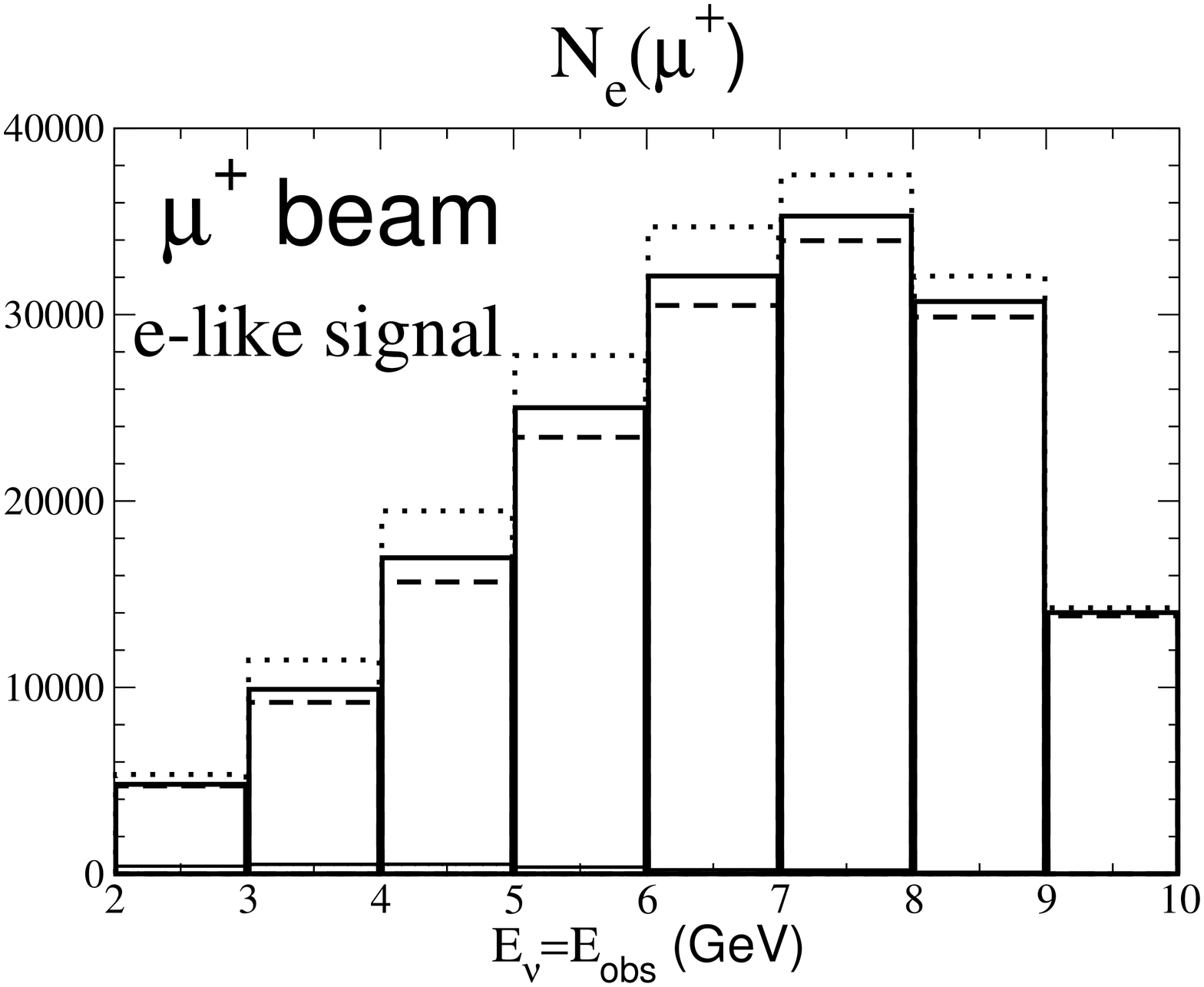,width=6cm,angle=-90}
\epsfig{file=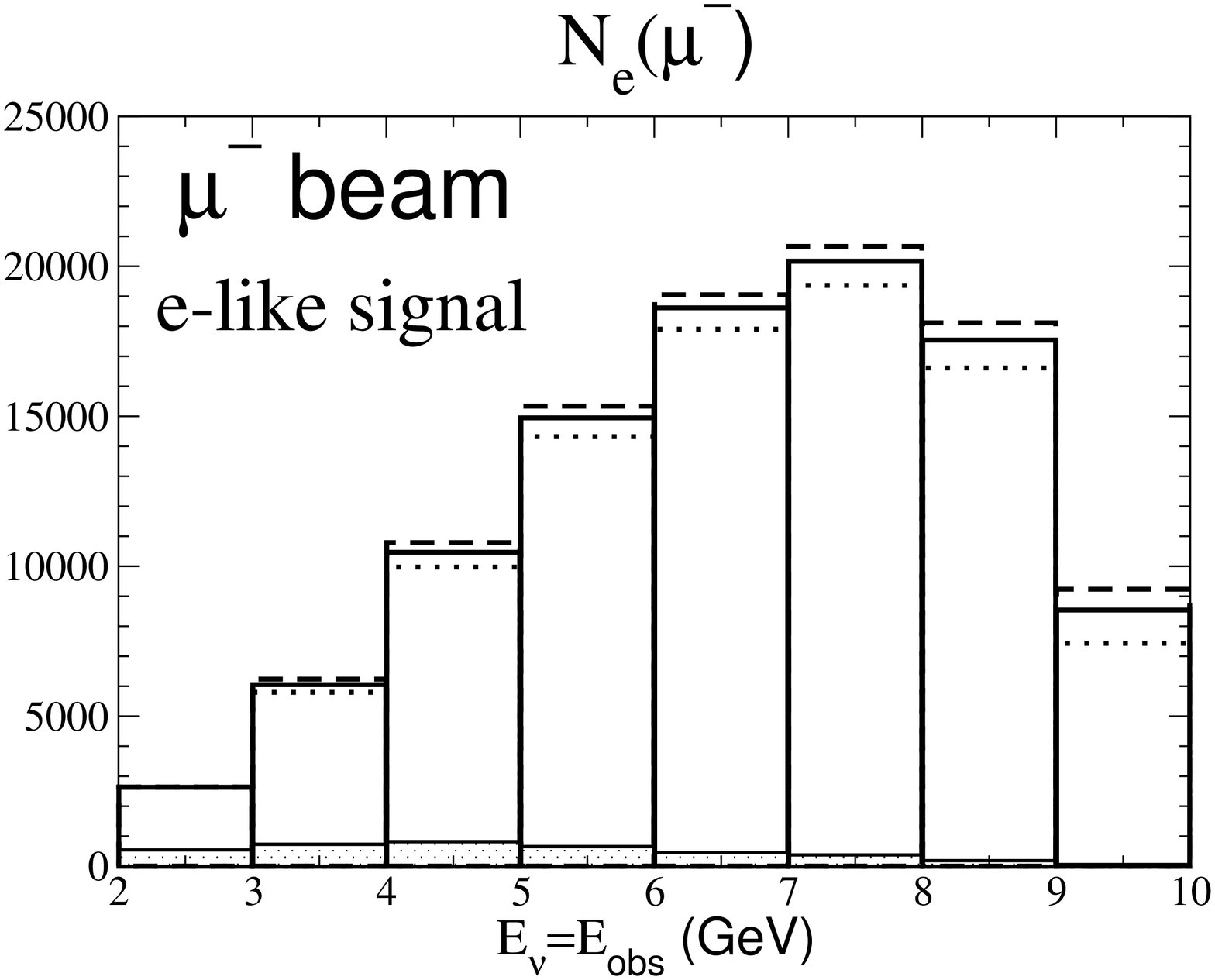,width=6cm,angle=-90}
\epsfig{file=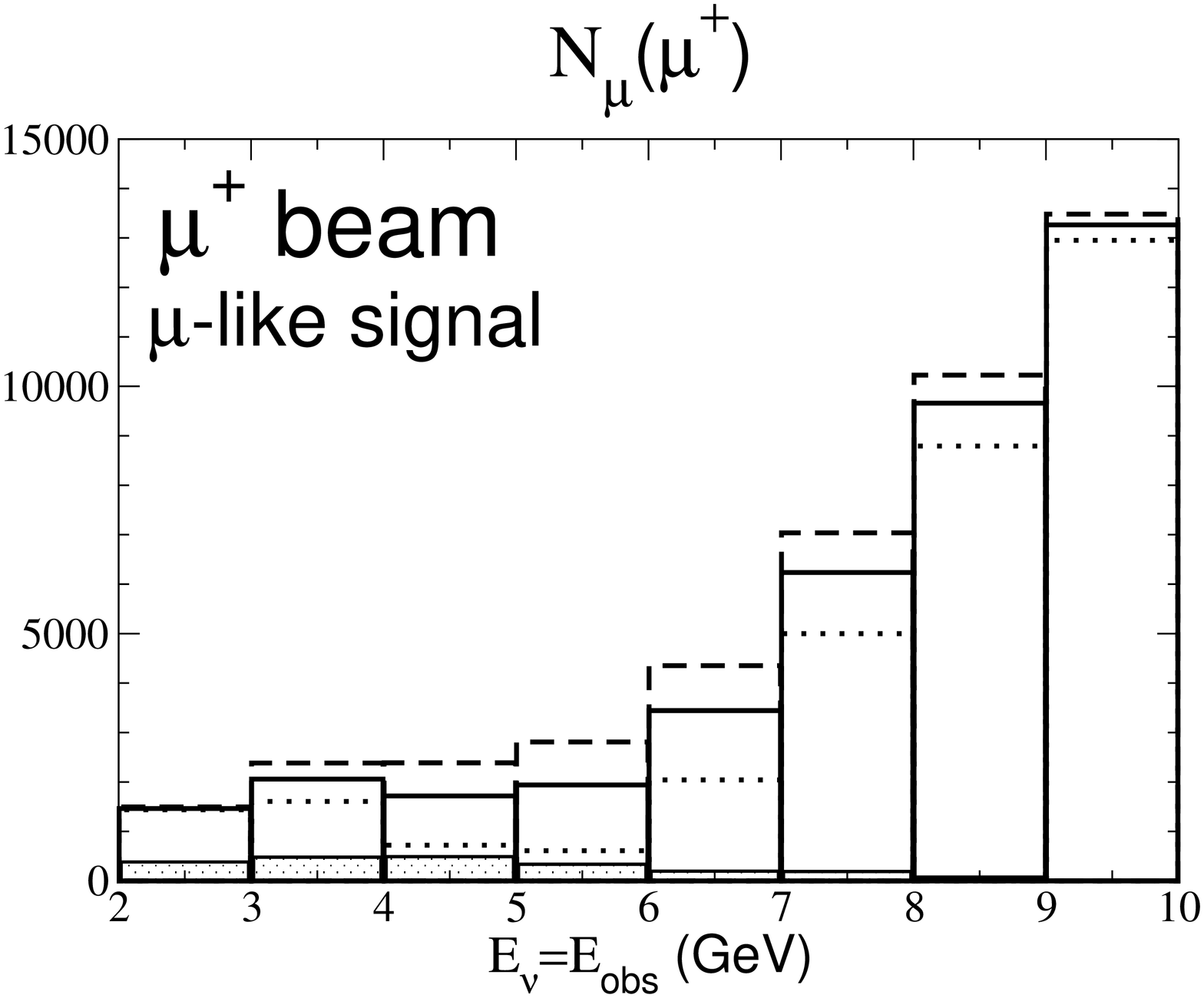,width=6cm,angle=-90}
\epsfig{file=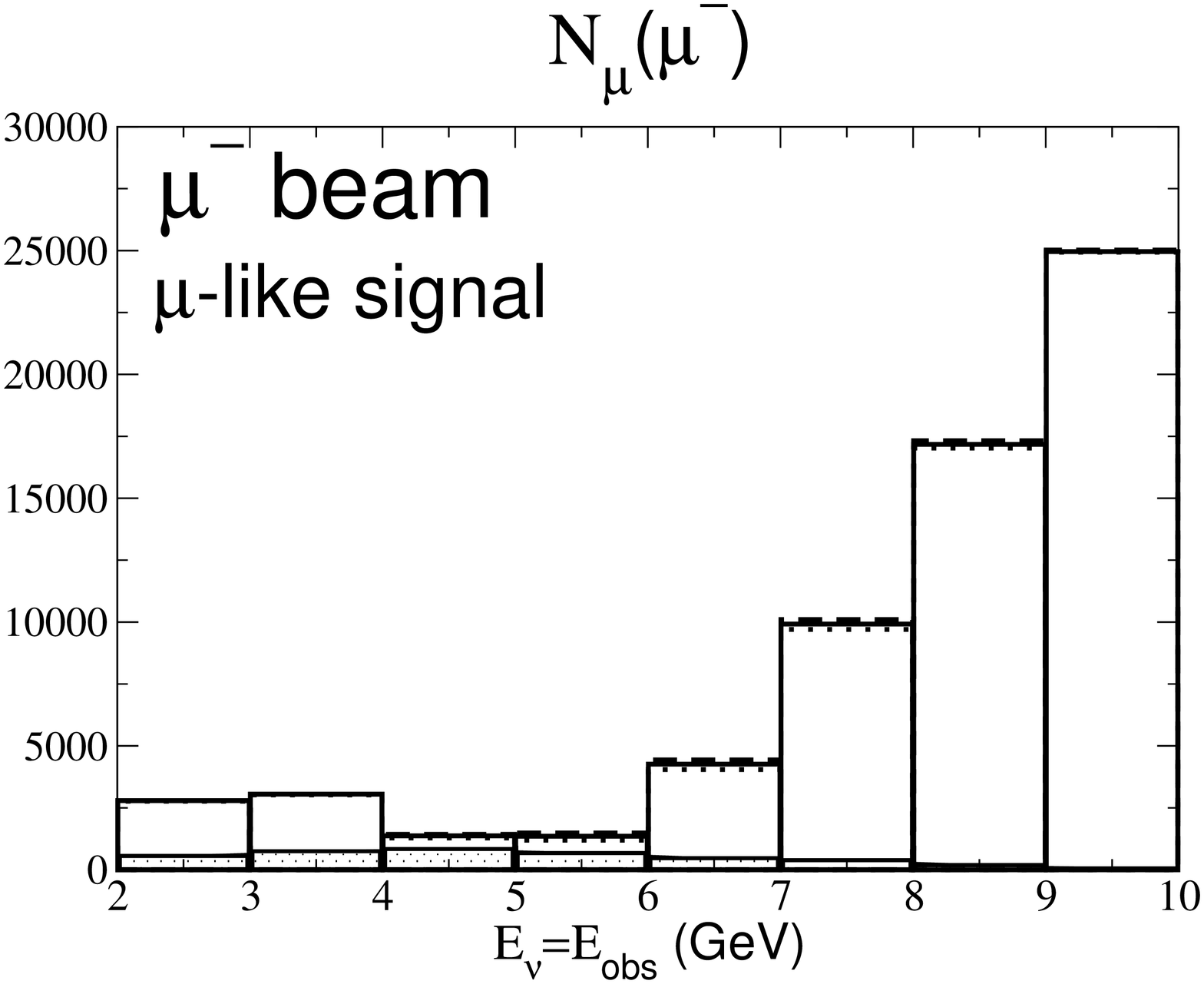,width=6cm,angle=-90}
\caption{
$\sin^22\theta_{_{\rm RCT}}$ dependences of the expected number of events.
The numbers are calculated for \eqref{signal_bin} but for 
$\sin^22\theta_{_{\rm RCT}}=0.1$ 
(dashed lines), 0.06 (solid lines), 0 (dotted lines) and
$\sin^2\theta_{_{\rm ATM}}=$0.5
with the normal-hierarchy.
Shaded bars show the number of the background events.
For the $\tau$ pureleptonic decay events, the horizontal 
scale is the observable energy after subtracting the final neutrino
energies.
}
\Fglab{signal_rct}
\end{center}
\end{figure}

The contribution from the survival mode dominates each signal but 
that in $N_\mu(\mu^+)$ and $N_\mu(\mu^-)$ vanish at
around $E_\nu \simeq 5$ GeV for $\sin^22\theta_{_{\rm ATM}}=1$
because of the nearly maximum oscillation.
In $N_e(\mu^+)$ and $N_\mu(\mu^-)$,
the contributions from the appearance mode,
$\ov \nu_\mu \to \ov \nu_e$ and 
$\ov \nu_e \to \ov \nu_\mu$, respectively,
are suppressed for the normal hierarchy,
because of the large matter effects, which suppresses 
$P_{\ov \nu_\mu \leftrightarrow \ov \nu_e}$
by a factor of 1/6 at $E_\nu\sim 5$ GeV \cite{H2B},
and also because of 
small anti-neutrino CC cross
sections,
$\sigma_{\ov \nu_l}/\sigma_{\nu_l}\simeq 1/2$.
Owing to these double suppressions of the anti-neutrino appearance 
contributions, $N_e(\mu^+)$ measures $P_{\nu_e\to \nu_e}$ and
$N_\mu(\mu^-)$ measures $P_{\nu_\mu\to \nu_\mu}$ in the normal
hierarchy. 
Because of this, $N_e(\mu^+)$ is very insensitive to the parameters
$\sin^2\theta_{_{\rm ATM}}$ and $\delta_{_{\rm MNS}}$ that affect 
the transition probabilities $P_{\nu_\mu \leftrightarrow \nu_e}$
and $P_{\ov \nu_\mu \leftrightarrow \ov \nu_e}$.
Insensitivity of $N_e(\mu^+)$ to $\sin^2\theta_{_{\rm ATM}}$ is
shown clearly in \Fgref{signal}.
The measurement of $N_e(\mu^+)$ hence gives us an opportunity 
to measure $\sin^22\theta_{_{\rm RCT}}$ uniquely in the normal 
hierarchy.
The $\sin^22\theta_{_{\rm RCT}}$ dependence of 
$N_e(\mu^+)$ is clearly seen in \Fgref{signal_rct}.

We find, on the other hand, that
the $\nu_e$ and $\ov \nu_e$ survival probabilities in the inverted
hierarchy are always larger and smaller than those in the normal hierarchy,
$P_{\nu_e \to \nu_e}^{\rm Norm} < P_{\nu_e \to \nu_e}^{\rm Inv}$ and
$P_{\ov\nu_e \to \ov\nu_e}^{\rm Norm} > P_{\ov\nu_e \to \ov\nu_e}^{\rm Inv}$,
because of the large matter effect.
In fact they are almost independent of $\sin^22\theta_{_{\rm RCT}}$ 
in the allowed region 
$0<\sin^22\theta_{_{\rm RCT}}\lsim 0.1$.
Because 
$P_{\nu_e \to \nu_e}^{\rm Norm} \sim P_{\nu_e \to \nu_e}^{\rm Inv}$
and
$P_{\ov\nu_e \to \ov\nu_e}^{\rm Norm} \sim P_{\ov\nu_e \to \ov\nu_e}^{\rm Inv}$
at $\sin^22\theta_{_{\rm RCT}}=0$, we find that the normal and inverted 
hierarchies can be distinguished by measuring $N_e(\mu^+)$ and $N_e(\mu^-)$, 
if $\sin^22\theta_{_{\rm RCT}}$ is not too small.
The differences are quite significant for 
$\sin^22\theta_{_{\rm RCT}}=0.06$, as shown in \Fgref{signal}.
If the degeneracy in the mass hierarchy is lifted,
the parameter $\sin^22\theta_{_{\rm RCT}}$ 
can be measured with good accuracy
by the observations of $N_e(\mu^+)$ and $N_e(\mu^-)$,
where the former is insensitive to 
$\sin^2\theta_{_{\rm ATM}}$ and $\delta_{_{\rm MNS}}$,
and so is the latter for very small $\sin^22\theta_{_{\rm RCT}}$.

The $\nu_\mu$ survival probability $P_{\nu_\mu \to \nu_\mu}$
depends on $\sin^22\theta_{_{\rm ATM}}$ 
in the leading term, which is measured uniquely by 
$N_\mu(\mu^-)$ for
the normal hierarchy owing to its small dependence on
$\sin^22\theta_{_{\rm RCT}}$ as shown in \Fgref{signal_rct}.
In \Fgref{signal}, the thick lines for the normal hierarchy show 
that the two values of $\sin^2\theta_{_{\rm ATM}}$ that give  
$\sin^2\theta_{_{\rm ATM}}=0.91$ give almost the same prediction.
This is because $P_{\ov\nu_e \to \ov\nu_\mu}$ is suppressed strongly
by the matter effect for the normal hierarchy.
On the other hand, in case of the inverted hierarchy,
we find that the degeneracy in $\sin^22\theta_{_{\rm ATM}}$ 
is resolved in $N_\mu(\mu^-)$.
This comes from the $\ov \nu_e \to \ov \nu_\mu$ mode 
which is enhanced by the matter effect and 
receives non-negligible contribution proportional to 
$\sin^2\theta_{_{\rm ATM}}\cdot\sin^2\theta_{_{\rm RCT}}$.

The signal of $N_\mu(\mu^+)$ is the sum of the CC events from $\nu_e\to\nu_\mu$
and $\ov\nu_\mu\to\ov\nu_\mu$, which depends on 
$\sin^2\theta_{_{\rm ATM}}\cdot\sin^2\theta_{_{\rm RCT}}$ and
$\sin^22\theta_{_{\rm ATM}}$, respectively. 
Both $\sin^2\theta_{_{\rm ATM}}$ and $\sin^22\theta_{_{\rm ATM}}$ 
contributions can be seen in \Fgref{signal}, and 
$\sin^22\theta_{_{\rm RCT}}$ dependence is clearly seen in 
\Fgref{signal_rct}.
Because of the nearly maximum oscillation,
the significant dependence on 
the sign of $\satms{}-1/2$ is seen at around $E_\nu \simeq 5$ GeV.
 
Finally, the $N_e(\mu^-)$ signal in \Fgref{signal} shows almost linear
dependence on 
$\sin^2\theta_{_{\rm ATM}}$ for the normal hierarchy, and that of  
\Fgref{signal_rct} shows a clear dependence on 
$\sin^22\theta_{_{\rm RCT}}$.
Those are the results of the
$\nu_\mu\to\nu_e$ transition mode which is enhanced 
by the matter effect in the normal hierarchy.

Summing up, we come up with the following strategy for resolving 
the degeneracies of the three neutrino model parameters if the 
neutrino mass hierarchy is normal.
If $\sin^22\theta_{_{\rm RCT}}$ is not too small, 
$N_e(\mu^+)$ 
and $N_e(\mu^-)$ 
determine the hierarchy, and $N_e(\mu^+)$ measures
$\sin^22\theta_{_{\rm RCT}}$.
$N_\mu(\mu^-)$ determines $\sin^22\theta_{_{\rm ATM}}$, and
both $N_\mu(\mu^+)$ and $N_e(\mu^-)$ are sensitive to 
$\sin^2\theta_{_{\rm ATM}}$.
We can foresee the sensitivity in $\delta_{_{\rm MNS}}$ as well,
because the last two observables are sensitive to 
$P_{\nu_\mu \leftrightarrow \nu_e}$.
The dependency in $\delta_{_{\rm MNS}}$ and $\pi -\delta_{_{\rm MNS}}$
may also be resolved, because their sum is sensitive to
$\cos\delta_{_{\rm MNS}}$ and their difference is sensitive to
$\sin\delta_{_{\rm MNS}}$.
A more complicated strategy is needed if the neutrino mass hierarchy 
turned out to be inverted. 
We will come back to this problem elsewhere.

Bellow, we perform an exploratory study of the potential of such 
experiments by assuming the following simple experimental setup:
A 100 kton water-$\check {\rm C}$erenkov calorimeter detector 
at $L=2,100$ km away from a neutrino factory, which delivers
$10^{21}$ decays of unpolarized
$\mu^+$ and $\mu^-$ at $E_\mu=10$ GeV.

We assume that the detector is capable of measuring the event energy,
$E_{\rm obs}=E_\nu({\rm beam})-E_\nu({\rm final})$,
with accuracy much better than 1 GeV in the interval 2 GeV 
$< E_{\rm obs} < 10$ GeV, and calculate the expected number of events
for 8 bins with the 1 GeV width.
The probability distribution of the three neutrino parameters is then
obtained from the following $\chi^2$ function:

\begin{eqnarray}
\chi^2 &=& 
\sum_{i=1}^8
\left\{
\left(
\dfrac{{N^{i}_{e}}(\mu^+)^{\rm fit} - 
       {N^{i}_{e}}(\mu^+)^{\rm obs}}
{\sqrt{N_e^i(\mu^+)^{\rm obs}}
}
\right)^2
+\left(
\dfrac{{N^{i}_{\mu}}(\mu^+)^{\rm fit} -
       {N^{i}_{\mu}}(\mu^+)^{\rm obs}}
{\sqrt{N_\mu^i(\mu^+)^{\rm obs}}
}\right)^2
\right\} \nn \\
&&+
\sum_{i=1}^8
\left\{\left(
\dfrac{{N_{e}^i}(\mu^-)^{\rm fit} - 
       {N_{e}^i}(\mu^-)^{\rm obs}}
{\sqrt{N_e^i(\mu^-)^{\rm obs}}
}\right)^2
+\left(
\dfrac{{N^{i}_{\mu}}(\mu^-)^{\rm fit} -
       {N^{i}_{\mu}}(\mu^-)^{\rm obs}}
{\sqrt{N_\mu^i(\mu^-)^{\rm obs}}
}\right)^2
\right\} \nn \\
&&
+
\left(
\dfrac{\msun{\rm fit}-\msun{\rm true}}{0.1\times \msun{\rm true}}
\right)^2
+
\left(
{\dfrac{\ssun{\rm fit}-\ssun{\rm true}}{0.06}}
\right)^2 \nn \\
&&
+
\left(
{\dfrac{\matm{\rm fit}-\matm{\rm true}}{10^{-4}}}
\right)^2
+
\left(
{\dfrac{\satmw{\rm fit}-\satmw{\rm true}}{0.01}}
\right)^2 
+
\left(
{\frac{\rho-3.0^{^{}}}{0.1_{_{}}}}
\right)^2 
\nn \\
&&
+\sum_{\alpha =e,\mu}
\left(
\frac{\varepsilon^D_\alpha-1^{^{}}}{\Delta\varepsilon^D_\alpha}
\right)^2 
+
\sum_{\beta =\nu,\ov \nu}
\left(
\frac{\varepsilon^\sigma_\beta-1^{^{}}}{\Delta\varepsilon^\sigma_\beta}
\right)^2 \nn
\\
&&
+\left(\frac{\varepsilon_\tau -1}{\Delta \varepsilon_\tau}\right)^2
+\left(\frac{\varepsilon_{_{e/{\rm NC}}} -1}{\Delta 
\varepsilon_{_{e/{\rm NC}}}}
\right)^2
\,. 
\eqlab{def_chi}
\end{eqnarray}
Here $N_l^i(\mu^\pm)^{\rm obs}$ are calculated as 
\bseq
\bea
N_e^i(\mu^\pm)^{\rm obs}&=&N_e^i(\mu^\pm)+N_e^i(\mu^\pm,\tau\to e)
+N_e^i(\mu^\pm,\tau\to{\rm had})+N_e^i(\mu^\pm,{\rm NC})\,, \\
N_\mu^i(\mu^\pm)^{\rm obs}&=&N_e^i(\mu^\pm)+N_\mu^i(\mu^\pm,\tau\to\mu)\,,
\eea
\eseq
including contributions for the pure-leptonic $\tau$ decays,
and the fake contributions to $N_e^i(\mu^\pm)$ where the NC events
or the $\nu_\tau$ CC events with hadronic $\tau$ decays are 
mistaken as $e^\pm$ events.
These background contributions are calculated as
\bseq
\bea
N_l^i(\mu^+,\tau \to l)&=&
MN_A\int_{E_i}^{E_i+\delta E}dE_{\rm obs} 
\int_{E_0}^{E_{\rm max}}dE_\nu
\nn \\
&& 
\left\{ 
\Phi_{\ov \nu_\mu}\cdot P_{\ov \nu_\mu \to \ov \nu_\tau}\cdot
\frac{d\sigma^{\rm CC}_{\ov \nu_\tau}}{dE_{\rm obs}}
+
\Phi_{\nu_e}\cdot P_{\nu_e \to \nu_\tau}\cdot
\frac{d\sigma^{\rm CC}_{\nu_\tau}}{dE_{\rm obs}}
\right\}
\cdot {\rm B}(\tau \to l)\,,
\\
N_l^i(\mu^+,\tau \to {\rm had})&=&
MN_A\int_{E_i}^{E_i+\delta E}dE_{\rm obs} 
\int_{E_0}^{E_{\rm max}}dE_\nu
\nn \\
\left\{ 
\Phi_{\ov \nu_\mu}
\cdot
\right.
&
P_{\ov \nu_\mu \to \ov \nu_\tau}&
\left.
\cdot
\frac{d\sigma^{\rm CC}_{\ov \nu_\tau}}{dE_{\rm obs}}
+
\Phi_{\nu_e}\cdot P_{\nu_e \to \nu_\tau}\cdot
\frac{d\sigma^{\rm CC}_{\nu_\tau}}{dE_{\rm obs}}
\right\}
\cdot {\rm B}(\tau \to {\rm had})\cdot P_{e/{\rm NC}}
\,, \\
N_e^i(\mu^+,{\rm NC})&=&
MN_A\int_{E_i}^{E_i+\delta E}dE_\nu 
\left\{ 
\Phi_{\ov \nu_\mu}\cdot \sigma_{\ov \nu}^{\rm NC}
+
\Phi_{\nu_e}\cdot \sigma_\nu^{\rm NC}
\right\}
\cdot P_{e/{\rm NC}}\,,
\eea
\eseq 
and likewise for the events from decaying $\mu^-$.
Here $E_{\rm obs}$ is calculated by subtracting the energies
carried away by neutrinos in $\tau$ decays.
The oscillation probabilities are calculated for 
a set of parameters of the three neutrino model labeled
as `true', in the normal hierarchy for a constant matter 
density of $\rho=3$ g/cm$^3$.
The statistical errors of the number of events in each  
bin are simply the square root of the observed number calculated as above.

In the calculation of the model predictions for these numbers,
labeled as $N_l(\mu^\pm)^{\rm fit}$,
we allow all the 6 neutrino model parameters and the matter 
density $\rho$ to vary freely under the constraints given 
in the third and the fourth lines of \eqref{def_chi}.
For $\delta m^2_{_{\rm SOL}}$ and $\sin^22\theta_{_{\rm SOL}}$, 
we set the 1$\sigma$ errors which will be achieved 
by the KamLAND experiment \cite{KamLAND}. 
The LBL neutrino oscillation experiment Tokai-to-SK
will determine $\delta m^2_{_{\rm ATM}}$  
and $\sin^22\theta_{_{\rm ATM}}$
with the accuracy
$\delta(\delta m^2_{_{\rm ATM}}) \lsim 10^{-4}$ and 
$\delta(\sin^22\theta_{_{\rm ATM}}) < 0.01$, respectively \cite{JHF2SK}.
We assume that the average matter density along the baseline will be
known with 0.1g/cm$^3$ accuracy.

Furthermore, the following systematic effects are taken into account
in our fit.
Although we can probably neglect the flux uncertainties in the neutrino
factory experiments, the observed numbers of events should depend on
the detection efficiencies, and the expected numbers of events suffer
from the uncertainties in the neutrino CC cross sections.
We introduce the detection efficiency parameter 
$\varepsilon^D_e$ and $\varepsilon^D_\mu$
separately for
`$e$'-like and `$\mu$'-like events, 
respectively.
For simplicity,
we set $\varepsilon^D_e=\varepsilon^D_\mu=1$ in calculating the
expected number of events, but allow for their errors
\bea
\varepsilon^D_e=1\pm \Delta \varepsilon^D_e,~~~~~
\varepsilon^D_\mu=1\pm \Delta \varepsilon^D_\mu\,.
\eqlab{error_detect}
\eea
Here we assume that the fiducial volume uncertainty can be accounted
for as a part of the efficiency uncertainties.
For the cross sections, we assume that the ratio 
$\sigma_{\nu_\mu}^{\rm CC}/\sigma_{\nu_e}^{\rm CC}$ and
$\sigma_{\ov\nu_\mu}^{\rm CC}/\sigma_{\ov\nu_e}^{\rm CC}$ 
will be determined accurately in the future and introduce 
the following uncertainty factors
\bea 
\frac{\sigma_{\nu_\mu}^{\rm CC}}{(\sigma_{\nu_\mu}^{\rm CC})_{\rm input}}
=\frac{\sigma_{\nu_e}^{\rm CC}}{(\sigma_{\nu_e}^{\rm CC})_{\rm input}}
=\varepsilon_\nu^\sigma=1\pm\Delta\varepsilon_\nu^\sigma\,,~~
\frac{\sigma_{\ov\nu_\mu}^{\rm CC}}{(\sigma_{\ov\nu_\mu}^{\rm CC})_{\rm input}}
=\frac{\sigma_{\ov\nu_e}^{\rm CC}}{(\sigma_{\ov\nu_e}^{\rm CC})_{\rm input}}
=\varepsilon_{\ov\nu}^\sigma=1\pm\Delta\varepsilon_{\ov\nu}^\sigma\,.
\eqlab{error_cross}
\eea
These systematic uncertainties \eqref{error_detect} and \eqref{error_cross} 
play important roles in determining the ultimate accuracy of the
experiments when the statistical errors get smaller, and appear in the 
fifth line of \eqref{def_chi}.
Below, we show our results for 
$\Delta\varepsilon_{e}^D=\Delta\varepsilon_{\mu}^D
=\Delta\varepsilon_{\nu}^\sigma=\Delta\varepsilon_{\ov\nu}^\sigma=0$ 
and 2 \%,
so that the impacts of reducing these systematic uncertainties
can be inferred.
  
We also take account of the systematic errors for
the uncertainty in 
the branching fractions of $\tau$ leptonic-decays as 
\bea
\frac{B(\tau \to e \ov \nu_e \nu_\tau)}{B(\tau \to e \ov \nu_e
\nu_\tau)_{\rm input}}=
\frac{B(\tau \to \mu \ov \nu_\mu\nu_\tau)}{B(\tau \to \mu \ov
\nu_\mu\nu_\tau)_{\rm input}}=
\varepsilon_{\tau}=1\pm \Delta\varepsilon_{\tau}\,,
\eea
with
$B(\tau \to e \ov \nu_e \nu_\tau)_{\rm input}=0.178$ and
$B(\tau \to \mu \ov \nu_\mu\nu_\tau)_{\rm input}=0.174$ \cite{PDG},
and $\Delta \varepsilon_\tau=0.1$.
For the NC events, we set the mean value of the
$e/{\rm NC}$ misidentification probability as
$(P_{e/{\rm NC}}^{})_{\rm input} = 0.25 \% $ by using the estimations from the
K2K experiment \cite{K2K}.
The uncertainty factors are then introduced as 
\bea
\frac{P_{e/{\rm NC}}}{(P_{e/{\rm NC}})_{\rm input}}=\varepsilon_{e/{\rm
NC}}=1\pm \Delta\varepsilon_{e/{\rm NC}}\,,
\eea
with $\Delta\varepsilon_{e/{\rm NC}}=0.1$.
Contributions from these systematics are accounted for as the last 
two terms of \eqref{def_chi}.

\begin{figure}[ht]
\begin{center}
\epsfig{file=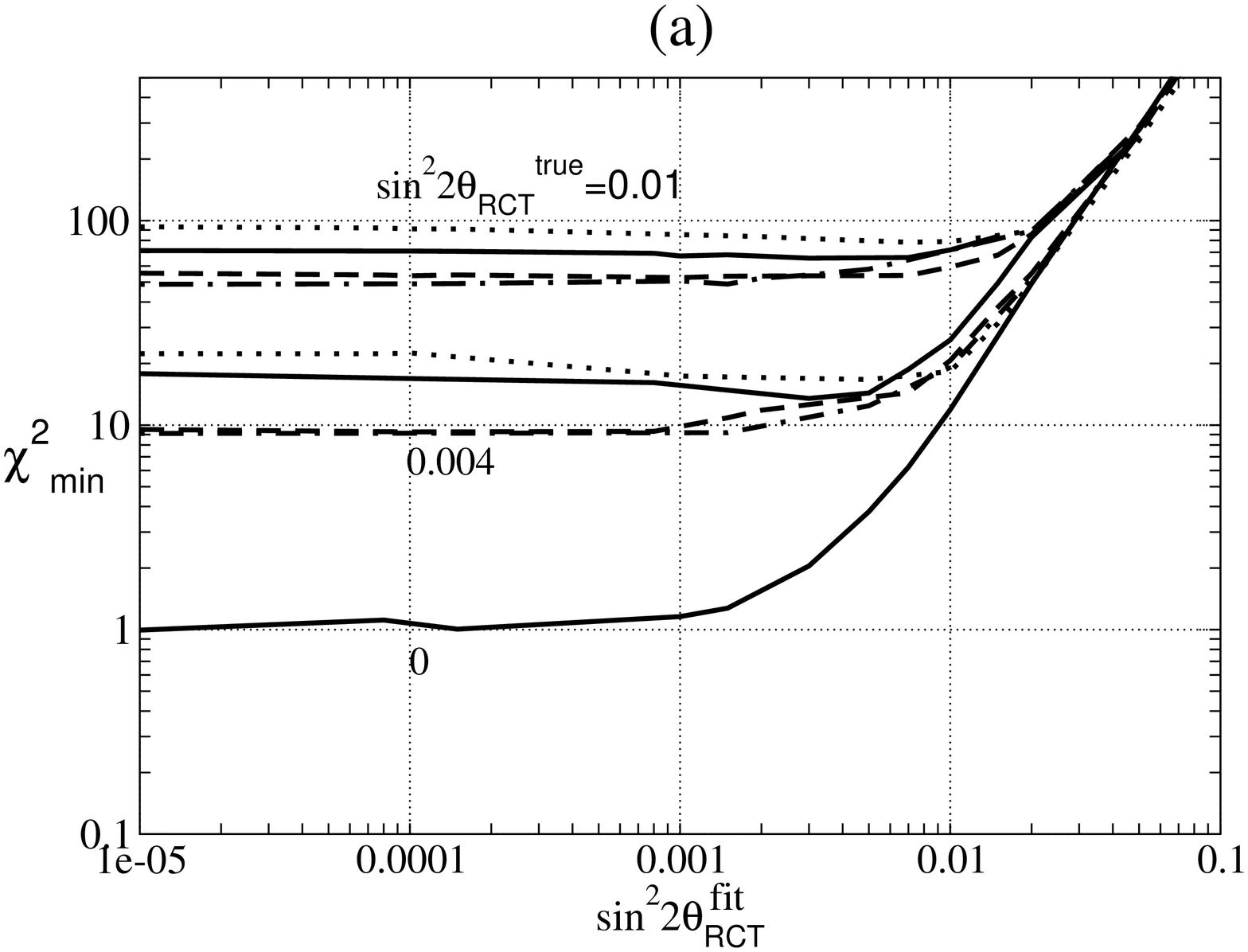,width=6.2cm,angle=-90}
\epsfig{file=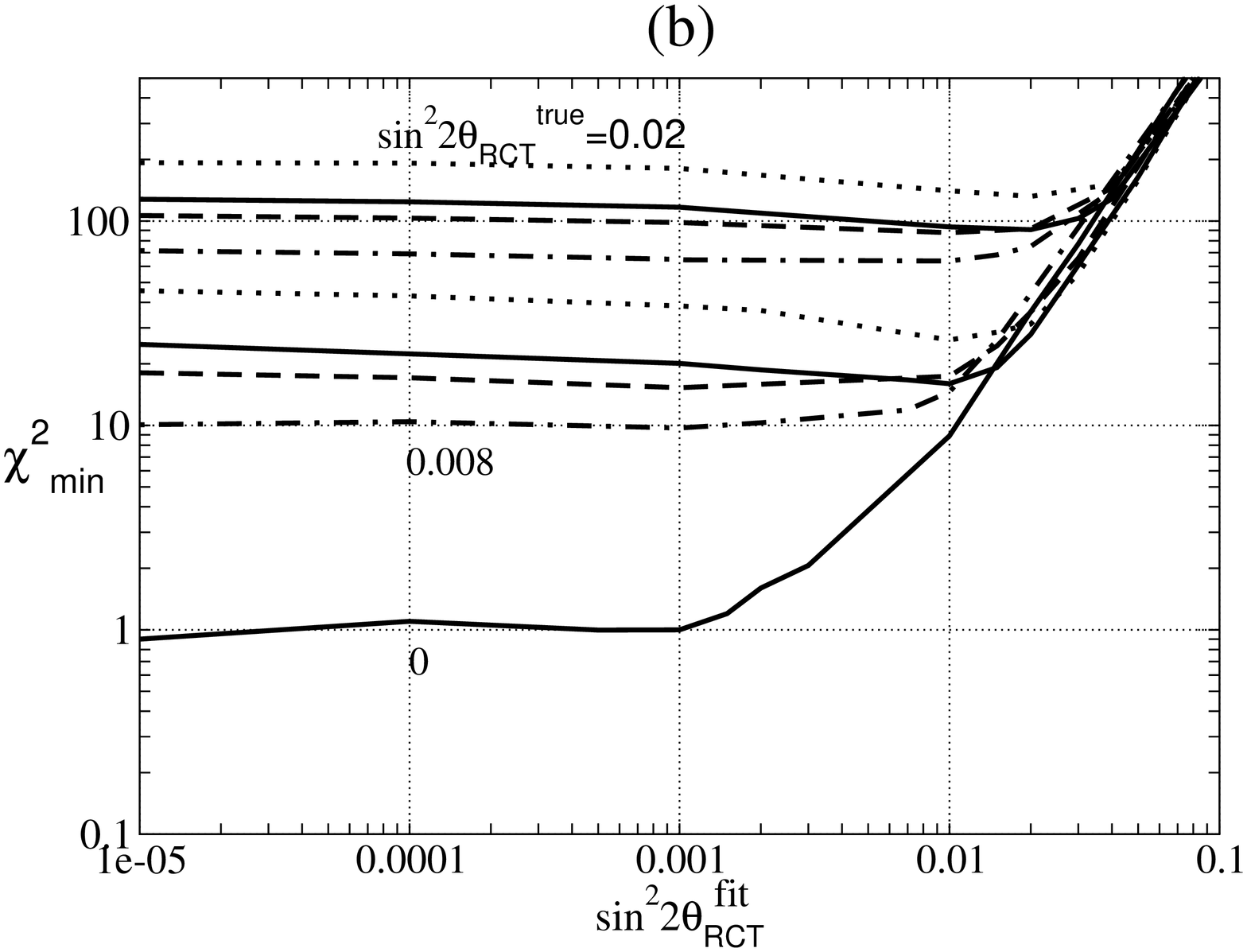,width=6.2cm,angle=-90}
\caption{%
$\chi^2_{\rm min}$ as a function of the fitting parameter of
 $\sin^22\theta_{_{\rm RCT}}^{\rm fit}$ by assuming the inverted hierarchy.
The resluts are for $10^{21}$ $\mu^+$ and $\mu^-$ decays for a 100 kton
water target at $L=2,100$ km.
The input data 
are calculated for the `true' values in
\eqref{true} by assuming the normal hierarchy
The 12 sets of the input data are labeled by 
$\sin^22\theta_{_{\rm RCT}}^{\rm true}=0.01\,,0.004\,$ and 0
for (a), 0.02\,,0.008\, and 0 for (b),
with
$\delta_{_{\rm MNS}}^{\rm true}= 0^\circ$ (solid lines), 
$90^\circ$ (dotted lines),~$180^\circ$ (dashed lines),
$270^\circ$ (dot-dashed lines).
$\Delta\varepsilon_{e}^D=\Delta\varepsilon_{\mu}^D
=\Delta\varepsilon_{\nu}^\sigma=\Delta\varepsilon_{\ov\nu}^\sigma=0$ 
for (a),
and
$\Delta\varepsilon_{e}^D=\Delta\varepsilon_{\mu}^D
=\Delta\varepsilon_{\nu}^\sigma=\Delta\varepsilon_{\ov\nu}^\sigma=0.02$
for (b). 
}
\Fglab{hierarchy}
\end{center}
\end{figure}

We show in \Fgref{hierarchy} $\chi^2_{\rm min}$ as a
function of $\sin^22\theta_{_{\rm RCT}}^{\rm fit}$ 
when the event numbers calculated for the normal hierarchy 
are analyzed by assuming
the inverted 
hierarchy. 
The results shown in \Fgref{hierarchy} (a)
are obtained when the detecting efficiencies and the cross sections are
known exactly,
$\Delta\varepsilon_{e}^D=\Delta\varepsilon_{\mu}^D
=\Delta\varepsilon_{\nu}^\sigma=\Delta\varepsilon_{\ov\nu}^\sigma=0$.
Because of 
$P_{\nu_e \to \nu_e}^{\rm Norm} < P_{\nu_e \to \nu_e}^{\rm Inv}$
and 
$P_{\ov\nu_e \to \ov\nu_e}^{\rm Norm} > P_{\ov\nu_e \to \ov\nu_e}^{\rm Inv}$,
the normal and inverted 
hierarchies can be distinguished by measuring $N_e(\mu^+)$
and $N_e(\mu^-)$. 
Normal hierarchy can be established at 5$\sigma$ level
when 
$\sin^22\theta_{_{\rm RCT}}^{\rm true}\gsim 0.01$,
and at 3$\sigma$ level when
$\sin^22\theta_{_{\rm RCT}}^{\rm true}\gsim 0.004$. 
However, the differences of the $e$-like signals 
between the normal and the inverted hierarchy is reduced to
$\sim 1\%$ when
$\sin^22\theta_{_{\rm RCT}}=0.004$. It is hence necessary to 
measure the cross section of $\nu_e$ and $\ov \nu_e$ CC events 
with an accuracy better than 1\%.
In \Fgref{hierarchy} (b), we show the results when we
account for 2 \% uncertainties in the cross sections and the efficiencies, 
$\Delta\varepsilon_e^D=\Delta\varepsilon_\mu^D
=\Delta\varepsilon_\nu^\sigma=\Delta\varepsilon_{\ov \nu}^\sigma=0.02$.
The hierarchy discrimination power of the experiment is reduced 
significantly, when the 5(3) $\sigma$ discrimination is now possible 
only when
$\sin^22\theta_{_{\rm RCT}}^{\rm true}\gsim 0.1 (0.008)$.

The input data for \Fgref{hierarchy} (a) and \Fgref{hierarchy} (b)
are generated by assuming the normal hierarchy
with the following `true' values :
\bseq
\eqlab{true}
\bea
\sin^2\theta_{_{\rm ATM}}^{\rm true}&=&0.5
\,, ~~~~~~~~~~~~~~~~~~~~~~~
\delta m^{2~{\rm true}}_{_{\rm ATM}}=3\times10^{-3} ~{\rm eV}^2
\,, 
\eqlab{true_atm}
\\
\sin^22\theta_{_{\rm SOL}}^{\rm true}&=&0.85
\,, ~~~~~~~~~~~~~~~~~~~~~~~
\delta m^{2~{\rm true}}_{_{\rm SOL}}=7\times10^{-5} ~{\rm eV}^2
\,, 
\eqlab{true_sol}
\\
\sin^22\theta_{_{\rm RCT}}^{\rm true}&=&0.01\,,~0.004\,,~0 ~({\rm for ~(a)})
\,,~~0.02\,,~0.008\,,~0 ~({\rm for~(b)})
\,,~~~~~
\\
\delta_{_{\rm MNS}}^{\rm true}&=&
 0^\circ,~90^\circ,~180^\circ,~ 270^\circ\, 
\eqlab{IvsIII_t_delta}\,.
\eea
\eseq
The 12 sets of the input data are labeled as
$\sin^22\theta_{_{\rm RCT}}^{\rm true}=0.01\,,0.004\,$ and 0
in \Fgref{hierarchy} (a) and $0.02\,,0.008\,$ and 0 in 
\Fgref{hierarchy} (b)
with $\delta_{_{\rm MNS}}^{\rm true}= 0^\circ$ (solid lines), 
$90^\circ$ (dotted lines),~$180^\circ$ (dashed lines),
$270^\circ$ (dot-dashed lines).
We then fit the data by assuming the inverted hierarchy
by allowing all the parameters to vary freely, and obtain the 
$\chi^2_{_{\rm min}}$ values plotted in the figures.

\begin{figure}[ht]
\begin{center}
\epsfig{file=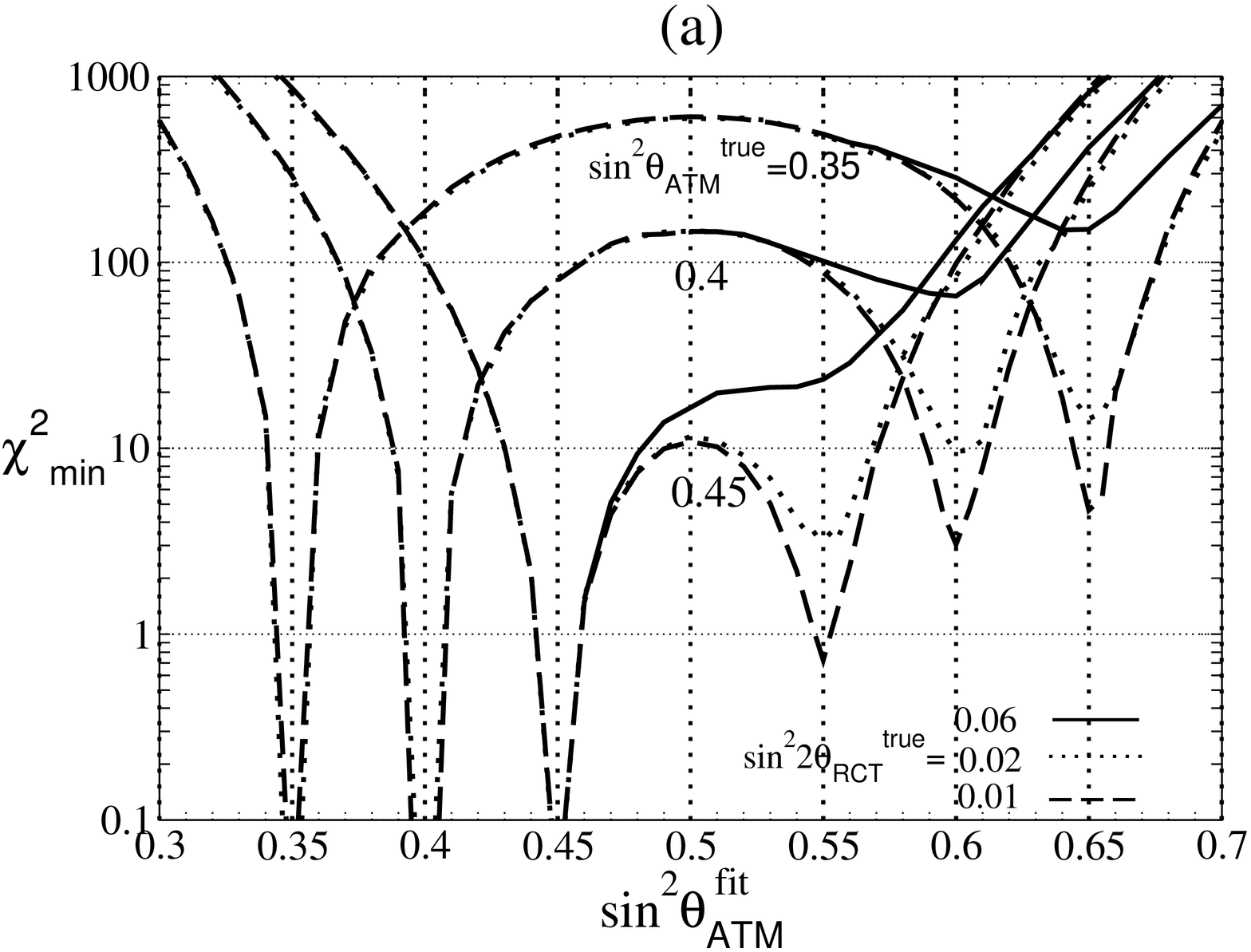,width=6cm,angle=-90}
\epsfig{file=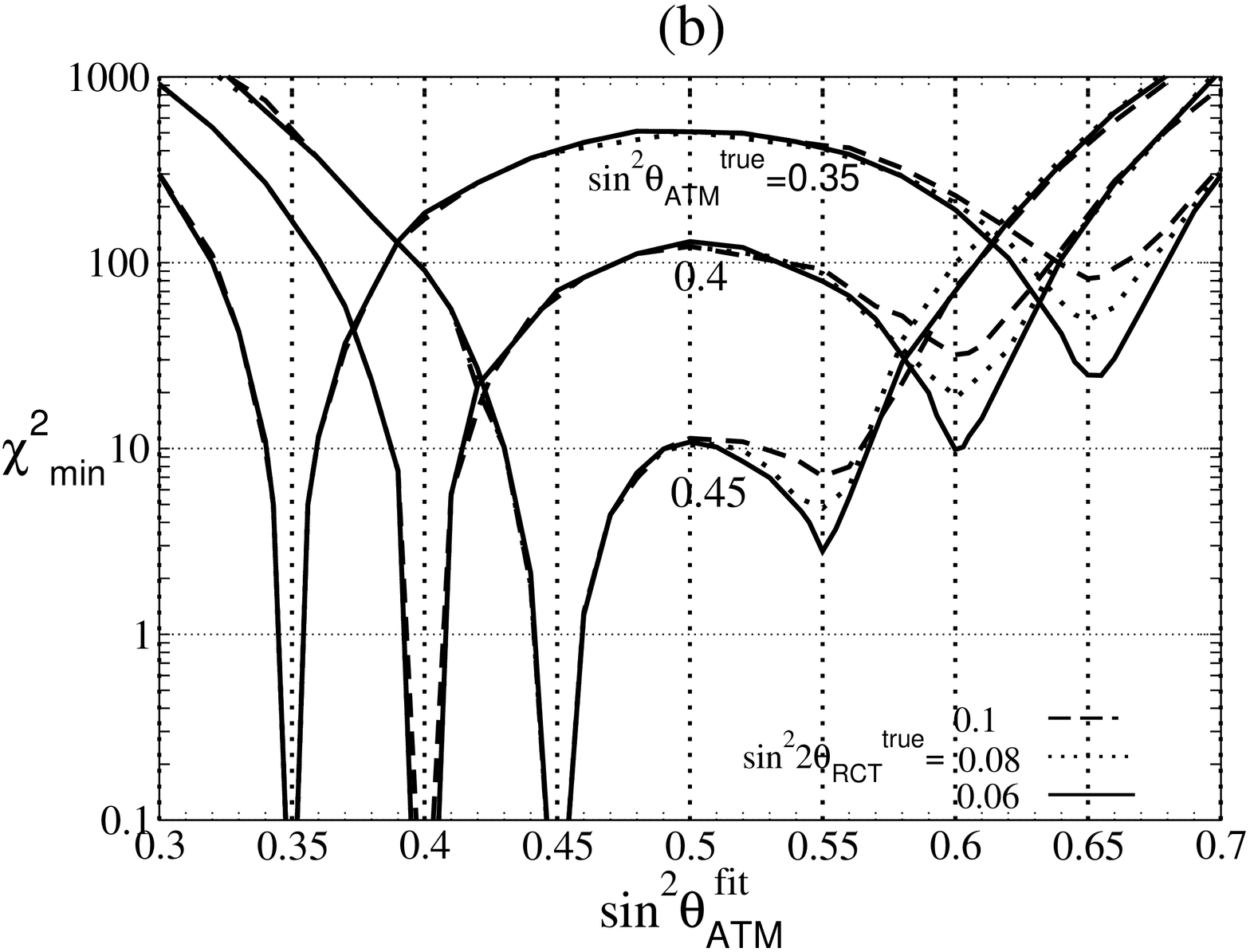,width=6cm,angle=-90}
\caption{%
$\chi^2_{\rm min}$ as a function of the fitting parameters 
$\sin^2\theta_{_{\rm ATM}}^{\rm fit}$
for (a) 
$\Delta\varepsilon_{e}^D=\Delta\varepsilon_{\mu}^D
=\Delta\varepsilon_{\nu}^\sigma=\Delta\varepsilon_{\ov\nu}^\sigma=0$ and
(b) 
$\Delta\varepsilon_{e}^D=\Delta\varepsilon_{\mu}^D
=\Delta\varepsilon_{\nu}^\sigma=\Delta\varepsilon_{\ov\nu}^\sigma=0.02$.
The input data 
of the events are calculated for 
$\sin^2\theta_{_{\rm ATM}}^{\rm true}=$0.35, 0.4, and 0.45 
with $\sin^22\theta_{_{\rm RCT}}^{\rm true}$ = 0.06~(solid lines),
0.02 [0.08] (dottted lines) and 0.01 [0.1] (dashed lines) for (a) [(b)]
and $\delta_{_{\rm MNS}}^{\rm true}=0^\circ$.
The other input values are the same as in
\eqref{true}.
}
\Fglab{atm}
\end{center}
\end{figure}
We show in \Fgref{atm} 
$\chi^2_{\rm min}$ as a function of
$\sin^2\theta_{_{\rm ATM}}^{\rm fit}$
for (a)
$\Delta\varepsilon_{e}^D=\Delta\varepsilon_{\mu}^D
=\Delta\varepsilon_{\nu}^\sigma=\Delta\varepsilon_{\ov\nu}^\sigma=0$, 
and 
(b) 
$\Delta\varepsilon_{e}^D=\Delta\varepsilon_{\mu}^D
=\Delta\varepsilon_{\nu}^\sigma=\Delta\varepsilon_{\ov\nu}^\sigma=0.02$. 
Input data
are calculated for 
$\sin^2\theta_{_{\rm ATM}}^{\rm true}=0.35\,$, 0.4\,, and 0.45\,
($\sin^22\theta_{_{\rm ATM}}^{\rm true}=0.91\,, 0.96\,$, and 0.99, 
respectively)
with three values of $\sin^22\theta_{_{\rm RCT}}^{\rm true} =0.06$
(solid lines), 0.02 [0.08] (dotted lines) and 0.01 [0.1] (dashed lines) 
for (a) [(b)]
and
$\delta_{_{\rm MNS}}^{\rm true}=0^\circ$ with the normal hierarchy.
The values of the other parameters are taken as in \eqref{true}.
The $\chi^2_{\rm min}$ function is found by varying
the fitting parameters
within the normal hierarchy.
We see that each $\chi^2_{\rm \min}$ has two dips at the value of
$\satms{\rm true}$ which give the same 
$\sin^22\theta_{_{\rm ATM}}$.
The results in \Fgref{atm} (a) show that we can resolve 
the degeneracy at 3$\sigma$ level 
when $\sin^2\theta_{_{\rm ATM}}^{\rm true}=0.4$ (0.45) for  
$\sin^22\theta_{_{\rm RCT}}^{\rm true} \gsim 0.02$ (0.04). 
The degeneracy in $\sin^2\theta_{_{\rm ATM}}$ can be resolved by 
the $\nu_e \to \nu_\mu$ mode in $N_\mu(\mu^+)$ and $\nu_\mu \to \nu_e$
mode in $N_e(\mu^-)$ whose leading terms are
proportional to $\sin^2\theta_{_{\rm ATM}}\sin^2\theta_{_{\rm RCT}}$,
when $\sin^22\theta_{_{\rm RCT}}$ and $\sin^22\theta_{_{\rm ATM}}$
are determined by the survival mode in $N_e(\mu^+)$ and
$N_\mu(\mu^-)$, respectively.
In \Fgref{atm} (b) where we take account of 2 \% systematic errors in 
the CC cross sections and the detection efficiencies, 
the degeneracy in the sign of 
$\sin^2\theta_{_{\rm ATM}}-1/2$ can be resolved at 3$\sigma$ level
when $\sin^2\theta_{_{\rm ATM}}^{\rm true}=0.4$ for  
$\sin^22\theta_{_{\rm RCT}}^{\rm true} \gsim 0.06$.
In the case of $\sin^2\theta_{_{\rm ATM}}^{\rm true}=0.45$, 
the determination of the sign is possible at 3$\sigma$ level 
only when 
$\sin^22\theta_{_{\rm RCT}}^{\rm true}\sim 0.1$.

In \Fgref{contour}, we show 
allowed regions in the plane of $\sin^22\theta_{_{\rm RCT}}^{\rm fit}$
and $\delta_{_{\rm MNS}}^{\rm fit}$
when $\sin^22\theta_{_{\rm RCT}}^{\rm true}=0.02\,, 0.06\,,$ and $0.1$
with
$\delta_{_{\rm MNS}} ^{\rm true}=0^{\circ}$ (upper left),
$90^{\circ}$ (upper right), $180^{\circ}$ (lower left), 
and $270^{\circ}$ (lower right)
in the normal hierarchy
and the other parameters are
\eqref{true_atm} and \eqref{true_sol}.
The errors in the efficiencies and the cross section are 
set as 
$\Delta\varepsilon_{e}^D=\Delta\varepsilon_{\mu}^D
=\Delta\varepsilon_{\nu}^\sigma=\Delta\varepsilon_{\ov\nu}^\sigma=0.02$. 
In each figure, the input parameter points are shown by solid-circles.
The normal hierarchy is assumed in the fitting.
The regions where $\chi^2_{\rm min} < 1$, 4, and 9 are depicted by
solid, dashed, and dotted boundaries, respectively.

\begin{figure}[ht]
\begin{center}
\epsfig{file=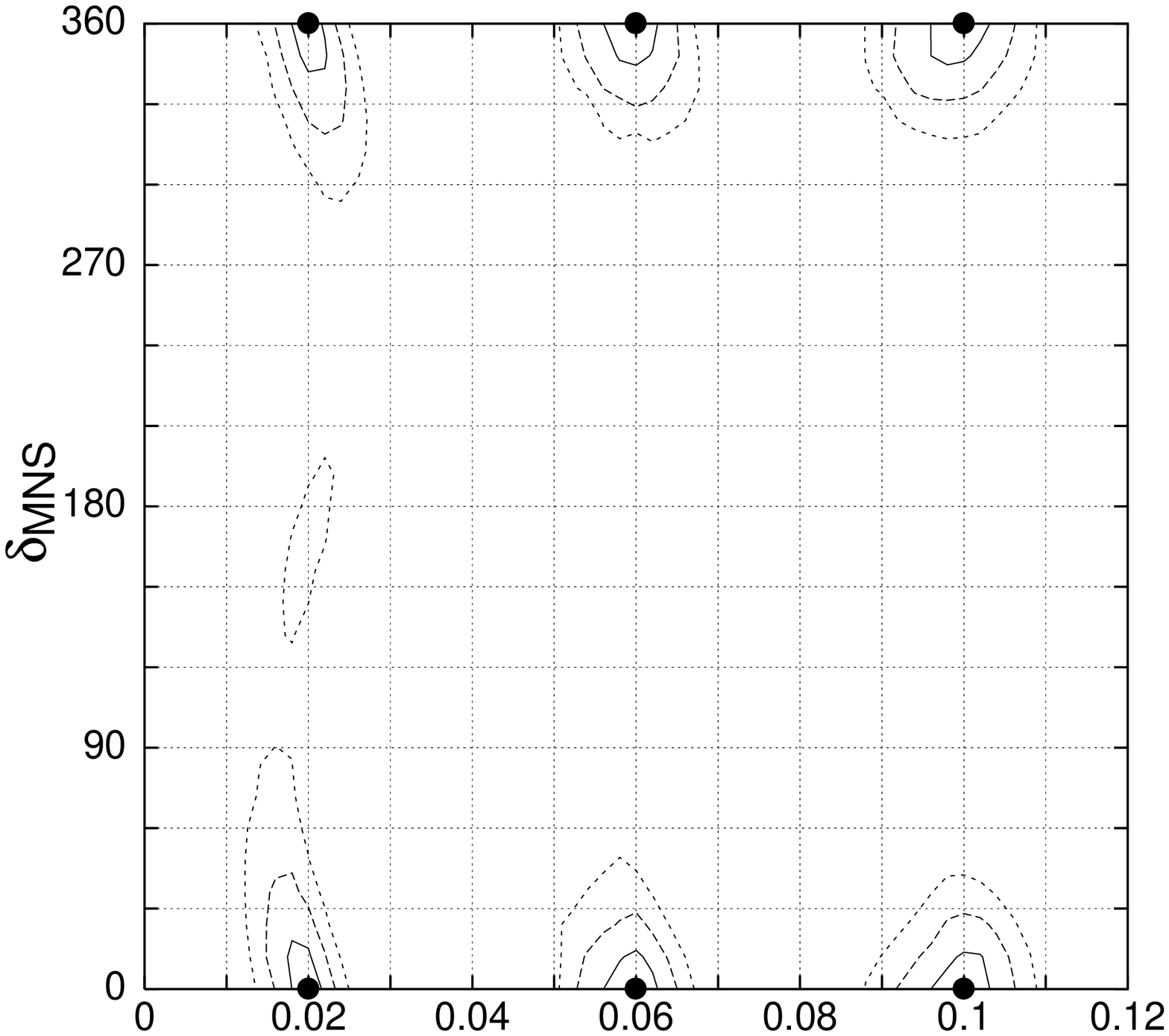,width=6cm,angle=-90}
\epsfig{file=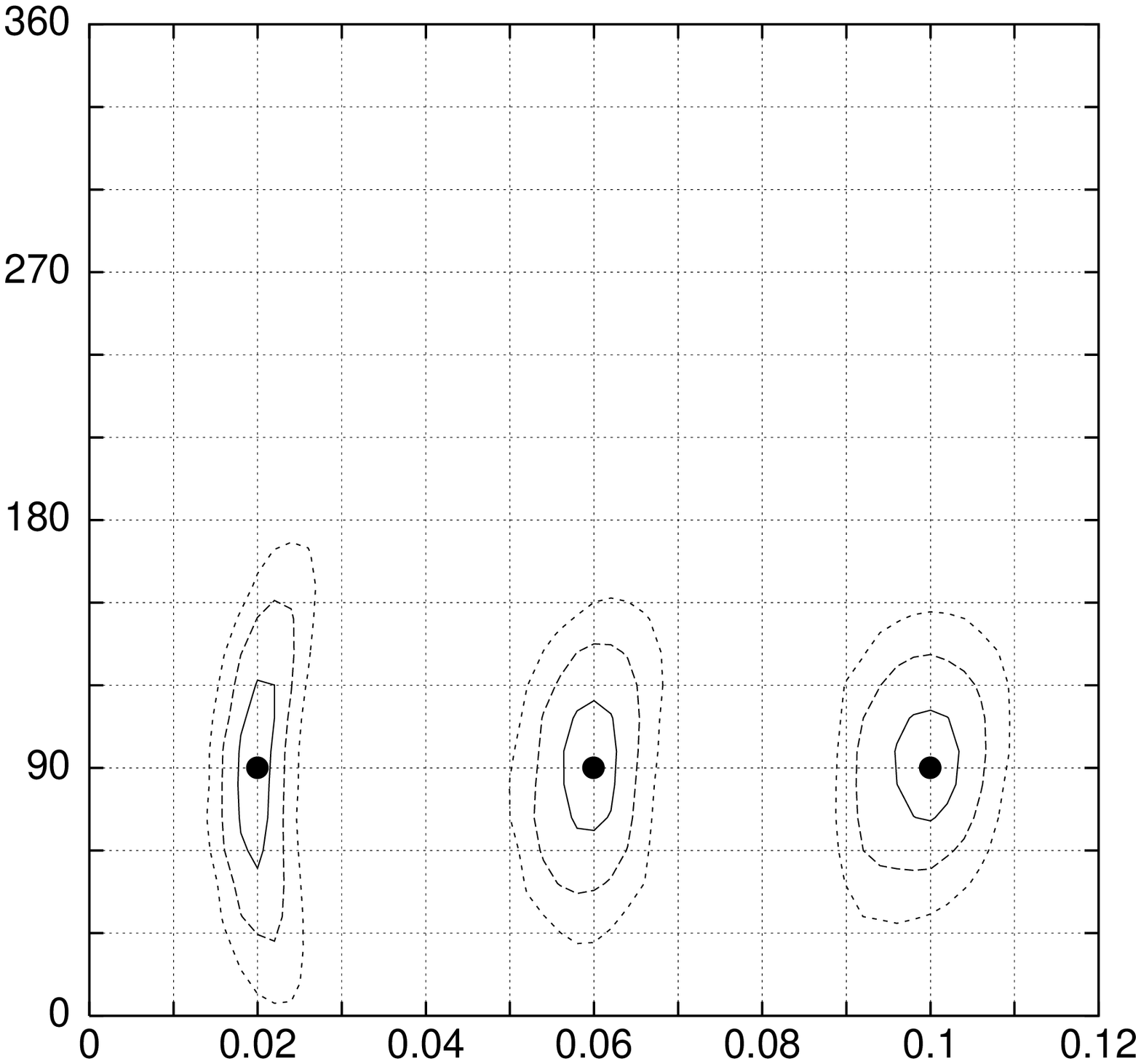,width=6cm,angle=-90}
\epsfig{file=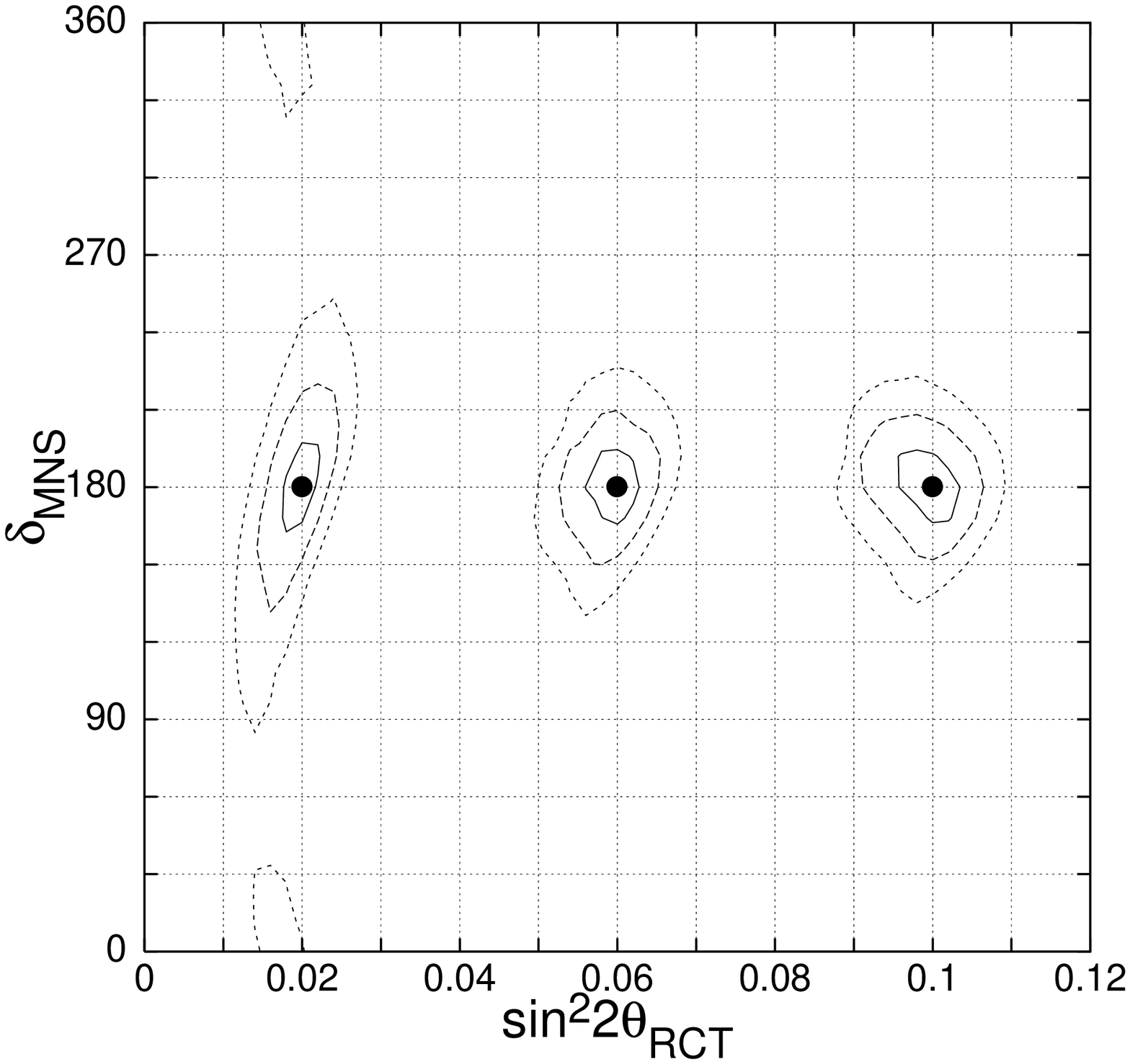,width=6cm,angle=-90}
\epsfig{file=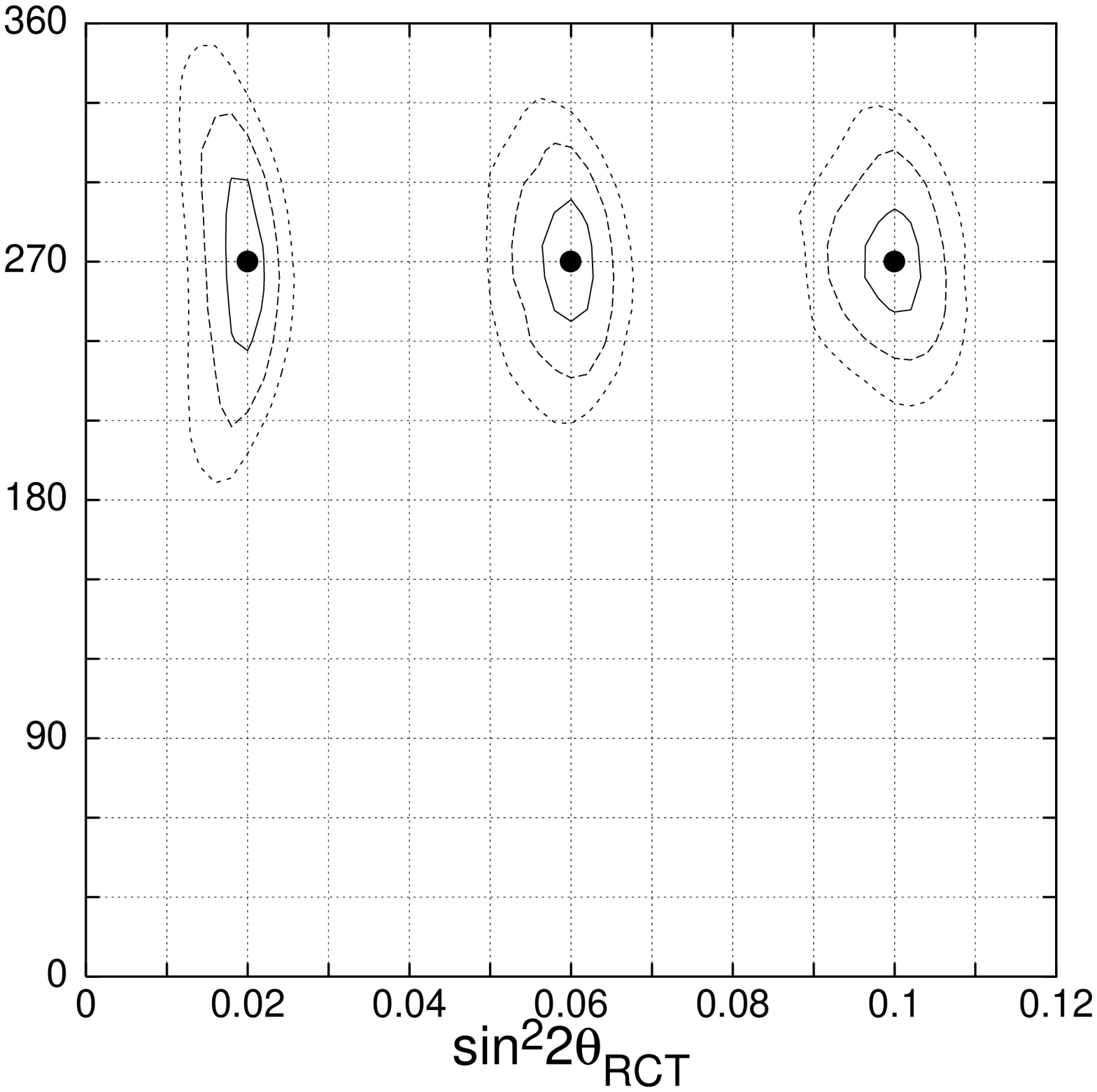,width=6cm,angle=-90}
\caption{%
Allowed regions in the plane of $\sin^22\theta_{_{\rm RCT}}$
and $\delta_{_{\rm MNS}}$
for $\sin^22\theta_{_{\rm RCT}}^{\rm true}=0.02\,, 0.06\,,$ and $0.1$
with
$\delta_{_{\rm MNS}} ^{\rm true}=0^{\circ}$ (upper left),
$90^{\circ}$ (upper right), $180^{\circ}$ (lower left), 
and $270^{\circ}$ (lower right).
The errors in the efficiencies and the cross section are 
taken to 
$\Delta\varepsilon_{e}^D=\Delta\varepsilon_{\mu}^D
=\Delta\varepsilon_{\nu}^\sigma=\Delta\varepsilon_{\ov\nu}^\sigma=0.02$. 
In each figure, the input parameter points are shown by solid-circles.
The regions where $\chi^2_{\rm min} < 1$, 4, and 9 are depicted by
solid, dashed, and dotted boundaries, respectively.
}
\Fglab{contour}
\end{center}
\end{figure}

The figures for
$\delta_{_{\rm MNS}}^{\rm true}=90^\circ$ and 
$\delta_{_{\rm MNS}}^{\rm true}=270^\circ$ show that 
the CP phase $\delta_{_{\rm MNS}}$ can be constrained locally
around the `true' points.
We can discriminate the maximal CP violation cases
$\delta_{_{\rm MNS}}^{\rm true}=90^\circ\,,270^\circ$ from
the CP conserving cases 
$\delta_{_{\rm MNS}}=0^\circ\,,180^\circ$ at 3$\sigma$ level
for $\sin^22\theta_{_{\rm RCT}}^{\rm true}\gsim 0.02$.
The sensitivity is based essentially on the T violating difference 
between $P_{\nu_e \to \nu_\mu}$ and $P_{\nu_\mu \to \nu_e}$. 
Because the matter effects enhance both of these transitions whereas
suppressed strongly the transitions $\ov \nu_\mu \to \ov \nu_e$ and
$\ov \nu_e\to \ov \nu_\mu$, 
$N_\mu(\mu^+)$ is sensitive to $P_{\nu_e \to \nu_\mu}$ and
$N_e(\mu^-)$ is sensitive to $P_{\nu_\mu \to \nu_e}$. 

For $\delta_{_{\rm MNS}}^{\rm true}=0^\circ$ and 
$\delta_{_{\rm MNS}}^{\rm true}=180^\circ$,
we can constrain $\delta_{_{\rm MNS}}$ within $\pm 45^\circ$ accuracy
for $\sin^22\theta_{_{\rm RCT}}^{\rm true}=0.06$ and 0.1,
while the fake solutions appear
for $\sin^22\theta_{_{\rm RCT}}^{\rm true}=0.02$ at 3$\sigma$ level.
This is essentially because both 
$\nu_e \to \nu_\mu$ and $\nu_\mu \to \nu_e$ 
transitions have terms proportional to 
$\sin^22\theta_{_{\rm RCT}}\cos\delta_{_{\rm MNS}}$; $N_e(\mu^+)$
measures $\sin^22\theta_{_{\rm RCT}}$ uniquely, and the sum of $N_\mu(\mu^+)$
and $N_e(\mu^-)$ is sensitive to
$\sin^22\theta_{_{\rm RCT}}\cos\delta_{_{\rm MNS}}$ while the difference
is sensitive to $\sin^22\theta_{_{\rm RCT}}\sin\delta_{_{\rm MNS}}$.
In our previous study of J-PARC superbeam and Hyper-Kamiokande \cite{H2H},
it was difficult to resolve the degeneracy which gives same values for
$\sin^22\theta_{_{\rm RCT}}\cos\delta_{_{\rm MNS}}$.

In this letter, we examine the capability of the VLBL
neutrino experiment with a neutrino factory at J-PARC in Tokai Village to 
resolve degeneracies in the neutrino oscillation parameters.
We find that a large segmented 
water-$\check{\rm C}$erenkov calorimeter detector
placed at a few thousand km away, such as a proposed BAND detector
\cite{BAND} in Beijing, can be very effective in solving the problem.
The charge identification capability at the detector is not required.
This is mainly because the $e^\pm$ events from the $\mu^+$ beam measure
the $\nu_e\to\nu_e$ survival rate rather accurately while the $e^\pm$
events from the $\mu^-$
beam are more sensitive to the $\nu_\mu\to\nu_e$ transition rate, 
for the normal hierarchy, where the earth matter effects strongly
suppress the $\ov \nu_\mu\to \ov\nu_e$ and $\ov \nu_e\to \ov\nu_\mu$
transitions. On the other hand, if the hierarchy is inverted, 
the $\nu_\mu\to \nu_e$ and $\nu_e\to\nu_\mu$ transitions are suppressed
and the $e^\pm$ events from $\mu^+$ beam are sensitive to 
the $\ov \nu_\mu\to \ov\nu_e$ transitions and those from $\mu^-$ beam
are sensitive to $\ov \nu_e\to \ov\nu_e$.
Because of the strong matter effects in the 
$\nu_\mu\leftrightarrow \nu_e$ and 
$\ov\nu_\mu\leftrightarrow\ov\nu_e$ transitions processes,
we can resolve not only the neutrino mass hierarchy but also 
the degeneracies in
the sign of $\sin^2\theta_{_{\rm ATM}}-1/2$ and those between
$\delta_{_{\rm MNS}}$ and $180^\circ-\delta_{_{\rm MNS}}$.

The following numerical results are found at 3 $\sigma$ level for
$10^{21}$ decaying unpolarized $\mu^+$ and $\mu^-$ at 10 GeV
when we take into account 
2\% uncertainties in the detection efficiencies 
and the $\nu_e$
and $\nu_\mu$ CC cross section measurements
and assume that the energy resolution is significantly smaller
than 1 GeV.
The mass hierarchy can be determined for 
$\sin^22\theta_{_{\rm RCT}}\gsim 0.008$,
which is remarkably smaller than the corresponding
limit in the VLBL experiment
with the conventional beams from J-PARC \cite{H2B}. 
When $\srct{}\gsim 0.06$,
the degeneracy in the sign of $\satms{}-1/2$
can be lifted for $\satmw{}=0.96$.
The CP violating phase 
$\delta_{_{\rm MNS}}$ can be uniquely constrained for
$\srct{}\gsim 0.02$ if its true value is around 
$90^\circ$ or $270^\circ$, while if it is
around $0^\circ$ or $180^\circ$,
the mirror solution at $180^\circ-\delta_{_{\rm MNS}}$ can be excluded
for $\srct{}\gsim 0.03$.
It is essential to have both $\mu^+$ and $\mu^-$ beams to obtain 
the above results, but it is not necessary for the detector to have
the charge identification capability.

{\it Acknowledgments}\\
The authors wish to thank stimulating discussions with Y.~Hayato,
T.~Kobayashi, M.~Koike, J.~Sato, Y.~F.~Wang and I.~Watanabe.
They would also like to thank the refree whose important
comments on the treatments of the systematic errors made our exploratory
study more realistic.
The work of MA is supported in part by the Academy of Finland 
and the Japan Society for the Promotion of Science (JSPS).
The work of KH has been partially supported by Japan-China Scientific 
Cooperation Program of JSPS.  
The work of NO is supported in part by a grand from 
the US Department of Energy, DE-FG05-92ER40709.

\end{document}